\documentclass[useAMS,usenatbib]{mn2e}
\usepackage{times}
\usepackage{epsfig}
\usepackage{graphicx}
\input epsf

\newcommand{\xmm}{{\it XMM~\/}}
\newcommand{\xmmn}{{\it XMM-Newton~\/}}
\newcommand{\asca}{{\it ASCA~\/}}
\newcommand{\chandra}{{\it Chandra~\/}}

\newcommand{\rosat}{{\it ROSAT~\/}}
\newcommand{\einstein}{{\it Einstein~\/}}

\def\ergcms{{\rm ~erg~cm^{-2}~s^{-1}}}

\def\ergsec{{\rm ~erg~s^{-1}}}
\def\cms{{\rm ~cm^{-2}}}

\def\H0{{\rm ~km~s^{-1}~Mpc^{-1}}}

\def\la{\mathrel{\hbox{\rlap{\hbox{\lower4pt\hbox{$\sim$}}}{\raise2pt\hbox{$<$}}}}}
\def\ga{\mathrel{\hbox{\rlap{\hbox{\lower4pt\hbox{$\sim$}}}{\raise2pt\hbox{$>$}}}}}
\def\d25{D$_{25}$}

\def\Ha{{H$\alpha$}}

\def\hii{H{\small II}$~$}

\def\deg{\hbox{$^\circ$ }}
\def\arcm{\hbox{$^\prime$ }}
\def\arcs{\hbox{$^{\prime\prime}$ }}

\title [\xmmn observations of NGC~7771/0 and NGC~2342/1] {\xmmn observations of the interacting galaxy pairs NGC~7771/0 and NGC~2342/1}
\author[L.P.\ Jenkins et al.]
	{L.P.\ Jenkins$^1$\thanks{E-mail: lej@star.le.ac.uk}, T.P.\ Roberts$^1$, M.J.\ Ward$^1$, A.\ Zezas$^2$\\ 
$^1$ X-ray \& Observational Astronomy Group, Dept. of Physics \& Astronomy, University of Leicester, University Road, Leicester LE1 7RH, U.K.\\
$^2$ Harvard-Smithsonian Center for Astrophysics, 60 Garden Street, Cambridge, MA 02138, USA. \\}

\date{Accepted ......................; Received .....................; in original form .....................}

\pagerange{\pageref{firstpage}--\pageref{lastpage}}
\pubyear{2004}

\begin{document}

\maketitle

\label{firstpage}

\begin{abstract}

We present \xmmn X-ray observations of the interacting galaxy pairs NGC~7771/7770 \& NGC~2342/2341. In NGC~7771, for the first time we are able to resolve the X-ray emission into a bright central source ($L_X\sim10^{41} \ergsec$), plus two bright ($L_X>10^{40} \ergsec$) ultraluminous X-ray sources (ULXs) located either end of the bar. The soft emission of the central source is well-modelled by a two-temperature thermal plasma with $kT$=0.4/0.7\,keV. The hard emission is modelled with a flat absorbed power-law ($\Gamma\sim1.7$, $N_H\sim10^{22} \cms$), and this together with a low-significance (1.7$\sigma$) $\sim300$\,eV equivalent width emission line at $\sim$6\,keV are the first indications that NGC~7771 may host a low-luminosity AGN. For the bar ULXs, a power-law fit to X-1 is improved at the 2.5$\sigma$ level with the addition of a thermal plasma component ($kT\sim0.3$\,keV), while X-2 is improved only at the 1.3$\sigma$ level with the addition of a disc blackbody component with T$_{in}\sim0.2$\,keV. Both sources are variable on short time-scales implying that their emission is dominated by single accreting X-ray binaries (XRBs). The three remaining galaxies, NGC~7770, NGC~2342 and NGC~2341, have observed X-ray luminosities of 0.2, 1.8 \& 0.9$\times10^{41} \ergsec$ respectively (0.3--10\,keV). Their integrated spectra are also well-modelled by multi-temperature thermal plasma components with $kT$=0.2--0.7\,keV, plus power-law continua with slopes of $\Gamma$=1.8--2.3 that are likely to represent the integrated emission of populations of XRBs as observed in other nearby merger systems. A comparison with other isolated, interacting and merging systems shows that all four galaxies follow the established correlations for starburst galaxies between X-ray, far-infrared and radio luminosities, demonstrating that their X-ray outputs are dominated by their starburst components.

\end{abstract}


\section{Introduction}
\label{sec:intro}

It is well known that starburst activity can be triggered, and possibly enhanced, by galactic encounters/collisions that may result finally in a merging system. The original N-body studies of \citet{toomre72} and many subsequent studies (e.g. \citealt{barnes96}) demonstrated how interactions can redistribute substantial quantities of material towards the central regions of the galaxies, triggering and fuelling violent bursts of star formation. Such encounters also create complex structures such as tidal tails, gas bridges and other disrupted morphologies, which are easily observable in the optical and radio wavelengths (e.g. \citealt{hibbard96}), and can influence the properties of all gas phases from cold neutral gas, to hot ionised gas.  Another possible consequence of tidal interactions is the formation of a galactic bar resulting from instabilities created in the disc (e.g. \citealt{noguchi88}; \citealt{barnes91}). Bars are believed to result in an inflow of gas to the central regions of the galaxy by the extraction of angular momentum from the gas in the disc through gravitational torques, thereby fuelling a starburst and/or an active galactic nucleus (AGN) \citep{barnes91}. 

Substantial observational evidence for enhanced star formation in interacting systems has been found at optical and infrared (IR) wavelengths (\citealt{larson78}; \citealt{joseph84}; \citealt*{lonsdale84}; \citealt{bushouse87}; \citealt*{bushouse88}; \citealt{kennicutt87};  \citealt{sanders96}). Although enhanced star formation activity is found in the discs of interacting galaxies, the effect is much stronger in the nuclear regions. Galaxies that are in close pairs or possess a bar show enhanced central radio emission, stronger than those of isolated/unbarred systems (e.g. \citealt{hummel90}), plus high nuclear emission-line luminosities and ionization levels (\citealt{kennicutt84}; \citealt{keel85}; \citealt*{bergvall03}). It is not currently clear whether galaxy-galaxy interactions are sufficient for triggering AGN activity, or whether the increased activity is entirely starburst in nature. For example, \citet{schmitt01} found that Seyfert galaxies have the same percentage of companion galaxies as other activity types (e.g. LINERs, \hii \& transition galaxies), though this could simply mean that we observe galaxies going through different stages of activity following the movement of gas to the central regions, as a delay is expected between the onset of an interaction (and an initial starburst phase), and the accretion of the material onto a nuclear black hole. Indeed, significant correlations between AGN activity and interactions are only found in high mass accretion rate systems such as very luminous/radio-loud quasars (see \citealt{jogee04} for a recent review).

X-ray studies offer a powerful means to understand the nature of starburst galaxies (SBGs), because enhanced X-ray emission is a manifestation of both star formation and AGN activity. With the launch of \rosat and {\it ASCA}, it became possible to observe the X-ray emission from interacting/merging galaxies (e.g. \citealt*{fabbiano97}; \citealt{read98}; \citealt{alonso99}; \citealt{henriksen99}).  However, the poor spatial resolution and relatively low effective area of these missions has hampered attempts to observe the X-ray morphology and detailed spectral properties of these systems. Now, with the combined spectral and imaging capabilities of \xmmn and {\it Chandra}, we have the opportunity to extend this work to a larger sample of systems at different stages of interaction. Many \xmmn and \chandra studies have been made in recent years of merging/closely interacting systems e.g. NGC~4438/39 (The Antennae, e.g. \citealt*{fabbiano01}), NGC~4485/4490 \citep{roberts02}, NGC~4676 (The Mice, \citealt{read03}) and post-merger galaxies e.g. NGC~3256 (\citealt{lira02}; \citealt{jenkins04b}) and Arp299 \citep*{zezas03}. Such studies have revealed that the hard (2--10\,keV) X-ray emission of SBGs are dominated by a handful of bright extra-nuclear point sources (see also \citealt{kilgard02}; \citealt{colbert04}), the positions of which generally correlate with regions of massive star formation. The brightest of these sources are the ultraluminous X-ray sources (ULXs), with X-ray luminosities $>10^{39} \ergsec$, exceeding the Eddington limit for accretion onto a 1.4$M_{\odot}$ neutron star by a factor of $\sim$ 10 (see section~\ref{sec:disc_ulx}).

Although many merging/post merger galaxies have been studied in detail, galaxies in earlier stages of interaction have recently been largely neglected. In this paper, we present \xmmn X-ray observations of NGC~7771/7770 and NGC~2342/2341, two pairs of SBGs that appear to be in an early stage of interaction. This paper is set out as follows. In section~\ref{sec:gals}, we summarise the multiwavelength properties of the two systems. In section~\ref{sec:data}, we outline our X-ray data reduction and analysis methods, and in section~\ref{sec:results} we present detailed results of the spectral fitting and timing analyses, comparing our results with previous X-ray observations of these systems. Finally, we discuss our results in section~\ref{sec:discuss} in the context of the properties of other interacting galaxies. Throughout this paper, we adopt distances to the galaxies assuming a value of $H_0$=75\,km\,s$^{-1}$\,Mpc$^{-1}$.

\begin{table*}
\caption{\xmmn observation details.}
 \centering
  \begin{tabular}{lccccccccc}
\hline
Target    & Observation ID & Date         & Filter & & \multicolumn{2}{c}{Useful exposure (s)} & & \multicolumn{2}{c}{Source count rate (ct s$^{-1}$, 0.3--10\,keV)}\\
          &                & (yyyy-mm-dd) &        & & PN            & MOS                     & & PN                 & MOS   \\ 
\hline
NGC~7771  & 0093190301     & 2002-06-21   & Medium & & 27148         & 31343                   & & 7.7$\times10^{-2}$ & 2.3$\times10^{-2}$ \\
X-1     &                &              &        & &               &                         & & 9.0$\times10^{-3}$ & 2.4$\times10^{-3}$ \\         
X-2     &                &              &        & &               &                         & & 8.1$\times10^{-3}$ & 1.8$\times10^{-3}$ \\
NGC~7770  &                &              &        & &               &                         & & 2.7$\times10^{-2}$ & 7.2$\times10^{-3}$ \\
\hline
NGC~2342  & 0093190501     & 2002-03-25   & Medium & & 24103         & 25096                   & & 8.7$\times10^{-2}$ & 2.7$\times10^{-2}$ \\
NGC~2341  &                &              &        & &               &                         & & 4.7$\times10^{-2}$ & 1.4$\times10^{-2}$ \\
\hline
\end{tabular}
\label{table:obs}
\end{table*}

\section{The Galaxies}
\label{sec:gals}

\subsection{NGC~7771/0}
\label{sec:n7771_intro}

NGC~7771 is a highly inclined ({\it i}$\sim$70$^{\circ}$) early-type barred spiral galaxy (SBa) at a distance of 56.7\,Mpc. It is part of the Lyon Group of Galaxies No. 483 consisting of NGC~7769, NGC~7771, NGC~7786, NGC~7770 and UGC~12828 \citep{garcia93}, and has also been shown to be interacting with a small edge-on disc galaxy NGC~7771A \citep{nordgren97}. NGC~7771, NGC~7770 and NGC~7769 are also catalogued as an interacting triplet \citep{karachentsev88}, although in this study we concentrate on the closely interacting pair NGC~7771/0, which have an angular separation of only $\sim$1.1 arcminutes on the sky corresponding to a projected physical separation of 18\,kpc at 56.7\,Mpc. Note that NGC~7769, located $\sim$ 5.4 arcminutes north-west of NGC7771, is detected as an extended X-ray source in this \xmm observation and will be included in an upcoming study. Of the other members of the group, NGC~7771A is not detected, and both NGC~7786 and UGC~12828 are outside the field-of-view of the observation.

NGC~7771 has a high far-IR (FIR) luminosity ($logL_{FIR}/L_\odot=11.1$, \citealt{sanders03}), classifying it as a luminous IR bright galaxy (LIRG). Previous studies in the optical, radio and IR show that NGC~7771 is going through a period of intense star formation (\citealt{smith99} and references therein). The nucleus has been classified as an H{\small II}-type from optical emission line ratios \citep{veilleux95}, and shows no evidence of broad \Ha\ emission expected from an AGN (\citealt{veilleux95}; \citealt*{davies97}). NGC~7771 has a disturbed morphology and a skewed \Ha\ rotation curve \citep{smith99}, both of which may be attributable to its interaction with NGC~7770. The central starburst region has been resolved at radio \citep{neff92} and near-IR (NIR) \citep{smith99} wavelengths into a nuclear source surrounded by a circumnuclear starburst ring $\sim$ 6 arcseconds (1.6\,kpc) in diameter. The ring is composed of radio-bright and NIR-bright clumps, with spatial displacements between the peaks of radio and NIR emission indicative of multiple generations of star formation spanning $\sim$4-10\,Myr \citep{smith99}.  NGC~7771 has been previously observed in X-rays using both the \einstein Observatory (\citealt*{fabbiano92}; \citealt*{rephaeli95}) and \rosat (\citealt{davies97}, see section~\ref{sec:n7771_comp}).

\subsection{NGC~2342/1}
\label{sec:n2342_intro}

NGC~2342/2341 is a more distant system at 71\,Mpc \citep{goldader97a}, and is included in the {\it Catalogue of Isolated Pairs of Galaxies in the Northern Hemisphere} \citep{karachentsev72}. The galaxies have an angular separation of $\sim$ 2.5 arcminutes, which corresponds to 52\,kpc at 71\,Mpc. This is not a particularly well studied pair, but both galaxies have high IR luminosities of $logL_{FIR}/L_\odot=10.8$ \citep{sanders03}. The largest galaxy of the pair is NGC~2342, classed as S(pec) \citep{devaucouleurs91}. It shows no evidence of AGN activity at any wavelength, and \citet*{ho97b} classify the nucleus as \hii rather than an AGN based on optical emission line ratios. This is also supported by the lack of broad \Ha\ emission expected from an AGN \citep{ho97c}, and the lack of a strong NIR continuum \citep{goldader97b}. This system was previously observed in the X-ray regime with the \rosat PSPC, and is included in the X-ray survey of galaxy pairs by {\citealt{henriksen99} (see section~\ref{sec:n2342_comp}).

\section{Observations, Data Reduction \& Analysis Techniques}
\label{sec:data}

\subsection{\xmmn Observations \& Reduction}

The details of the \xmmn observations are shown in Table~\ref{table:obs}. NGC~7771/0 was observed with \xmmn for $\sim$31\,ks on the 21st June, 2002; NGC~2342/1 for $\sim$29\,ks on the 25th March, 2002. Both observations were performed with the EPIC MOS-1, MOS-2 \& PN cameras in ``Prime Full Window'' mode with medium filters.  The event lists were pipeline-processed using the {\small SAS} ({\it Science Analysis Software}) {\small v5.3.1/5.3.2} for NGC~7771/0 and NGC~2342/1 respectively, and all data products (spectra, images \& lightcurves) were created using {\small SAS} {\small v5.4.1} using EPIC calibration as of February 2003 (NGC~7771) and September 2003 (NGC~2342). The data were filtered using standard patterns corresponding to single and double pixel events for the PN (0 \& 1--4), and patterns 0--12 (single to quadruple events) for the MOS, together with {\tt flag=0} to remove hot pixels and out-of-field events.

Both observations were checked for periods of high background soft proton flares by accumulating full-field light curves. There was no flaring present in the NGC~7771/0 observation, hence all $\sim$31\,ks of MOS-1 \& MOS-2 and $\sim$27\,ks of PN data were used in the subsequent analysis.  A small amount of flaring was present in the NGC~2342/1 observation, and the times of three peaks (each with count rates $>$ 50 counts per second in the 0.3--10 \,keV PN data) were screened from subsequent data analysis, leaving net good times of$\sim$25\,ks (MOS-1 \& MOS-2) and $\sim$24\,ks (PN).

\begin{figure}
\centering
\scalebox{0.7}{\includegraphics{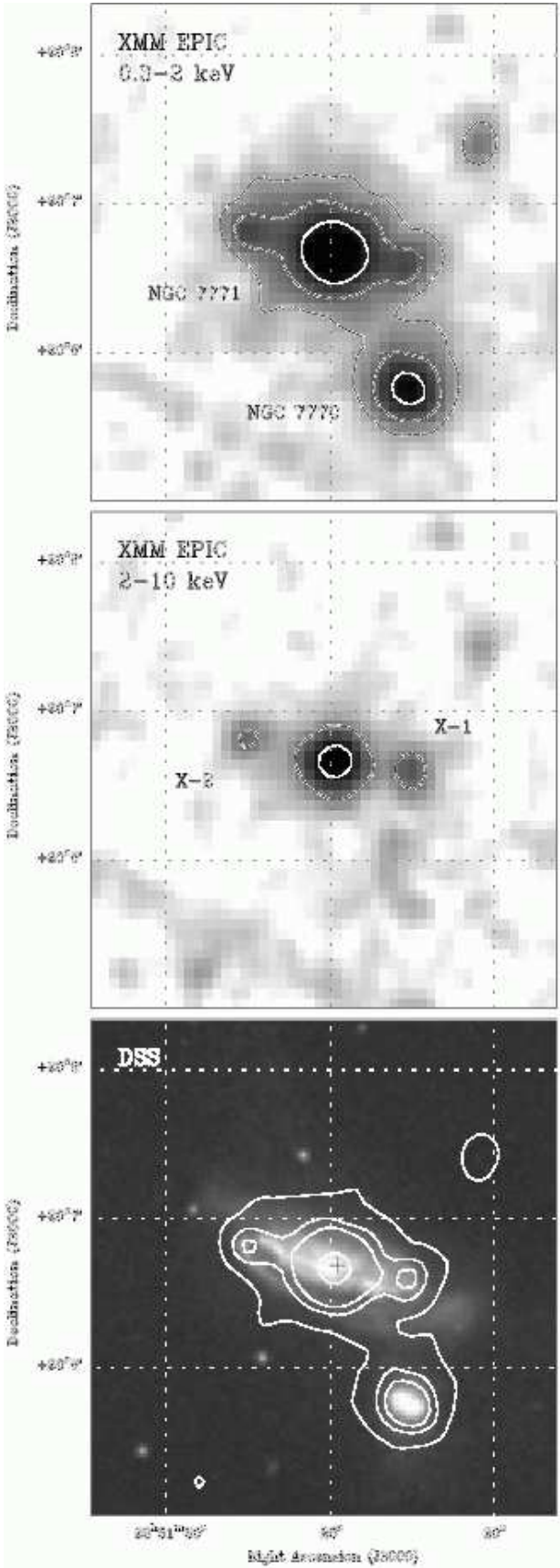}}
\caption{\xmmn EPIC soft (0.3--2\,keV, top) \& hard (2--10\,keV, middle) X-ray images of NGC~7771/0 with surface brightness contours overlaid. Bottom: DSS optical blue image of NGC~7771/0 with broad-band (0.3--10\,keV) X-ray contours overlaid. The cross marks the optical centre of the galaxy. See text for further details.}
\label{fig:n7771}
\end{figure}

\subsection{\xmmn Data Analysis}

We used the {\small SAS} task {\small ESPECGET} to simultaneously extract source and background spectra as well as create response matrices (RMFs) and ancillary response files (ARFs) for each source. The spectra for NGC~7771 and NGC~7770 were extracted using circular regions of 25 arcseconds radius, while the two discrete sources X-1 \& X-2 were extracted using circular regions of 12.5 arcseconds radius each. These regions were chosen to include the maximum emission from each source while avoiding contamination from nearby sources, and all sources were treated as point sources so that point-spread function (PSF) encircled energy corrections were applied. In each case, the background regions were taken close to each source, using as large an area as possible with approximately the same DET-Y distance as the source region (to ensure similar low-energy noise subtraction). For NGC~2342/1, the spectra were extracted using circular regions of 35 and 30 arcseconds radius respectively, with background regions selected in the manner just described. The resulting spectra were binned to a minimum of 20 counts per bin in order to optimise the data for $\chi^2$ statistics.

The spectral analysis of the \xmm data in this study has been performed using {\small XSPEC v11.1/11.3}. All errors are given at the 90 per cent confidence level unless stated otherwise.  The MOS and PN spectra have been fitted simultaneously in the 0.3-10\,keV band for each source, and we have included a free normalization constant to account for differences in the flux calibration of the three EPIC cameras, which differ by $\la$ 15 percent.

The spectral models we use to fit the data of the central source in NGC~7771 and the integrated emission from NGC~7770, NGC~2342 and NGC~2341 are a power-law continuum ({\small PL}) to represent the combined emission from compact sources, and an optically-thin thermal plasma ({\small MEKAL}, also denoted {\small M}) to model the thermal components expected in SBGs such as galactic winds and supernova remnants (SNRs). These types of models have been used successfully to describe the broad-band X-ray emission detected in other SBGs with \asca (NGC~253 \& M82; \citealt{ptak97}), and {\it XMM-Newton/Chandra} (e.g. NGC~253, \citealt{pietsch01}, \citealt{strickland02}; M82, \citealt*{stevens03}; NGC~3256 \& NGC~3310, \citealt{lira02}, \citealt{jenkins04b}). For this study we have chosen to fix the abundance parameters in the {\small MEKAL} models to solar values, as previous results have shown that fitting complicated multi-temperature thermal plasmas with this type of simplistic spectral model tends to result in unrealistically low metal abundances of $\sim$0.05--0.3Z${\odot}$ (\citealt{stricklandstevens00} and references therein). For the discrete sources in NGC~7771, we have also applied models typically used to model emission from X-ray binaries (XRBs) such as the multicolour disc blackbody ({\small DISKBB}) model, which describes thermal emission from an an accretion disc (\citealt{mitsuda84}; \citealt{makishima86}) and a conventional blackbody ({\small BBODY}) model.

\section{Results}
\label{sec:results}

\subsection{NGC~7771/0}
\label{sec:n7771_results}

\subsubsection{X-ray morphology}
\label{sec:n7771_morph}

Figure~\ref{fig:n7771} shows soft (0.3--2\,keV, top) and hard (2--10\,keV, middle) band \xmmn EPIC images of the the NGC~7771/0 system. The images are the sum of all 3 detectors, and are convolved with a 1$\sigma$ (4 arcsecond) 2-D Gaussian mask and displayed with a logarithmic scale. The intensity minima/maxima correspond to surface brightnesses of $1.2\times10^{-5}$/$4.4\times10^{-4}$ ct s$^{-1}$ (soft) and $8.8\times10^{-6}$/$3.5\times10^{-4}$ ct s$^{-1}$ (hard). The contours correspond to 0.4,1.1 \& $2.8\times10^{-4}$ ct s$^{-1}$ (soft) and 0.6 \& $2.8\times10^{-4}$ ct s$^{-1}$ (hard). Figure~\ref{fig:n7771} (bottom) shows a DSS blue image of NGC~7771/0, overlaid with broad-band (0.3--10\,keV) X-ray contours corresponding to surface brightnesses of 0.6,1.4,2.2 \& 9.4$\times10^{-4}$ ct s$^{-1}$.

The soft X-ray emission is prevalent throughout the central and bar regions of NGC~7771. For the first time, we are able to resolve the hard X-ray emission into a bright central source plus two sources located west and east of the central region. The DSS image (Figure~\ref{fig:n7771}, bottom) shows that the X-ray emission of the central source correlates well with the optical centre of NGC~7771. The best available (J2000) position for the centre of NGC~7771 is $\alpha$=23h 51m 24.8s, $\delta$=+20d 06m 42.6s $\pm1.25$ arcseconds from the 2MASS All-Sky Catalog \citep{cutri03}, and is marked with a cross on the OM image. This agrees to the centroid of the X-ray emission ($\alpha$=23h 51m 24.7s, $\delta$=+20d 06m 42.2s) to within the \xmm positional error of $\pm1.5$ arcseconds (which includes both statistical and systematic pointing errors). 

The two off-nuclear sources in NGC~7771 are officially designated XMMU J235122.3+200637 (west) and XMMU J235127.2+200652 (east), although in this paper they are referred to as X-1 and X-2 respectively. They are located at either end of the galactic bar at distances of $\sim$34 arcseconds (X-1) and 36 arcseconds (X-2) from the nucleus, which places them almost exactly at the corotation radius of the bar (35 arcseconds, \citealt{smith99}). We have checked whether the central and bar sources are extended by fitting their radial profiles in the stacked (PN+MOS) images using the {\small IRAF} task {\it imexam}, and comparing their full-widths at half-maximum (FWHM) to that of the \xmm EPIC PSF ($\sim$ 6 arcseconds). Both the nuclear source and the bar source X-1 show evidence of extended emission, especially in the soft 0.3--2\,keV band where the FWHM for each measures $\sim$10--11 arcseconds. X-2 may also be slightly extended with a FWHM of $\sim$ 8 arcseconds. 

The soft (0.3--2\,keV) X-ray emission from the dwarf companion NGC~7770 correlates well with the optical image, although it does not appear to be a strong source of hard (2--10\,keV) emission (this agrees with its steep spectral shape, see section~\ref{sec:n7770_spec}). There is evidence that the soft emission is extended, with a FWHM of up to 12 arcseconds in the soft (0.3--2\,keV) EPIC images. In addition, there appears to be low-level soft diffuse X-ray component extending from the south of NGC~7771 towards NGC~7770 that could be a consequence of tidal interactions between the galaxies. The lowest surface brightness contour that traces this feature in Figure~\ref{fig:n7771} (top left) represents flux level $\sim$ 5 times that of the average background flux level in that area. However, the distance between the peak of the emission in both galaxies is $\sim$ 1 arcminute, and we cannot rule out that this is an artifact of the large PSF of the EPIC instruments ($\sim45$ arcsecond 90 percent encircled energy radius at 1.5\,keV). A deep, high-resolution \chandra observation is required to ascertain whether this feature is real.

\subsubsection{NGC~7771 Spectral Analysis}
\label{sec:n7771_spec}

\begin{table*}
\caption{Spectral fitting results for NGC~7771/7770.}
 \centering
  \begin{tabular}{lccccccccccc}
\hline
\multicolumn{2}{c}{{\small PL}} & & \multicolumn{2}{c}{M$_1$} & & \multicolumn{2}{c}{M$_2$}   &  $\chi^2$/dof  & $F_X$$^b$     & \multicolumn{2}{c}{$L_X$$^c$}   \\ 

\\[-3mm]

$N_H$$^a$              & $\Gamma$  & & $N_H$$^a$ & $kT (keV)$    & & $N_H$$^a$  & $kT (keV)$  & &              &  Obs            & Unabs         \\
\hline
\multicolumn{12}{c}{\bf NGC~7771 Central Source}\\
\hline

\multicolumn{12}{l}{{\bf Model 1:} {\small M}+{\small PL} {\small (WABS*(PO+MEKAL))}}\\

1.07$^{+0.20}_{-0.21}$ & 1.45$^{+0.06}_{-0.08}$ & & $\dag$ & 0.51$^{+0.06}_{-0.05}$ & & - & - &  152.3/156 & 2.81$^{+0.18}_{-0.25}$ & 1.08$^{+0.07}_{-0.10}$ & 1.28$^{+0.07}_{-0.14}$ \\

\\[-2mm]
\multicolumn{12}{l}{{\bf Model 2:} {\small M}+{\small PL} {\small (WABS*PO+WABS*MEKAL)}}\\

1.39$^{+0.79}_{-0.59}$  & 1.49$^{+0.12}_{-0.13}$ & & $<$1.01                & 0.53$^{+0.08}_{-0.05}$ & & - & -  & 151.5/155 & 2.80$^{+0.11}_{-0.34}$ & 1.08$^{+0.04}_{-0.13}$ & 1.25$^{+0.02}_{-0.21}$ \\

\\[-2mm]
\multicolumn{12}{l}{{\bf Model 3: {\small M}+{\small M}+{\small PL} {\small (WABS*PO+WABS*MEKAL+WABS*MEKAL)}}}\\

{\bf 10.17$^{+7.74}_{-8.09}$} & {\bf 1.66$^{+0.13}_{-0.28}$} & & {\bf $<$0.43} & {\bf 0.35$^{+0.05}_{-0.06}$} & & {\bf 6.57$^{+2.50}_{-2.24}$} & {\bf 0.67$^{+0.16}_{-0.09}$}  & {\bf 141.1/152} & {\bf 2.78$^{+0.30}_{-0.85}$} & {\bf 1.07$^{+0.11}_{-0.33}$} & {\bf 2.24$^{+0.03}_{-1.04}$} \\

\hline
\multicolumn{12}{c}{\bf NGC~7770}\\

\hline
\multicolumn{12}{l}{{\bf Model 1:} {\small M}+{\small PL} {\small (WABS*(PO+MEKAL))}}\\

0.72$^{+0.43}_{-0.45}$ & 2.41$^{+0.42}_{-0.32}$ & & $\dag$ & 0.50$^{+0.07}_{-0.14}$ & & - & -  & 52.3/55 & 0.59$^{+0.06}_{-0.13}$ & 0.23$^{+0.02}_{-0.05}$ & 0.31$^{+0.03}_{-0.08}$ \\

\\[-2mm]
\multicolumn{12}{l}{{\bf Model 2:} {\small M}+{\small PL} {\small (WABS*PO+WABS*MEKAL)}}\\

$<$1.18 & 2.16$^{+0.23}_{-0.39}$  & & 3.40$^{+2.10}_{-1.47}$ & 0.30$^{+0.13}_{-0.09}$ & & - & - & 49.7/54 & 0.61$^{+0.04}_{-0.21}$ & 0.23$^{+0.01}_{-0.08}$ & 0.59$^{+0.02}_{-0.40}$ \\

\\[-2mm]
\multicolumn{12}{l}{{\bf Model 3: {\small M}+{\small M}+{\small PL} {\small ({\small WABS}*(PO+MEKAL+MEKAL))}}}\\

{\bf 1.36$^{+1.47}_{-0.82}$} & {\bf 2.26$^{+0.45}_{-0.40}$} & & $\dag$ & {\bf 0.15($<$0.23)} & & $\dag$ & {\bf 0.51($<$0.60)}  & {\bf 44.2/53} & {\bf 0.60$^{+0.06}_{-0.16}$} & {\bf 0.23$^{+0.02}_{-0.06}$} & {\bf 0.42$^{+0.03}_{-0.18}$} \\

\hline
\end{tabular}
\begin{tabular}{l}
Notes: ~Spectral models: {\small M}={\small MEKAL} thermal plasma (solar abundances) and {\small PL}=power-law continuum model. ~$^a$Absorption column in \\units of $10^{21}$ cm$^{-2}$. ~$^b${\it Observed} fluxes in the 0.3--10\,keV band, in units of $10^{-13}$ erg s$^{-1}$ cm$^{-2}$. ~$^c${\it Observed} and {\it unabsorbed} luminosities in \\the 0.3--10\,keV band, in units of $10^{41} \ergsec$ (assuming a distance of 56.7\,Mpc). ~$\dag$Same hydrogen column as applied to the previous \\spectral component. Errors correspond to 90 per cent confidence limits for parameter of interest. The best-fit models are highlighted in bold.\\
\end{tabular}
\label{table:n7771}
\end{table*}

\begin{figure*}
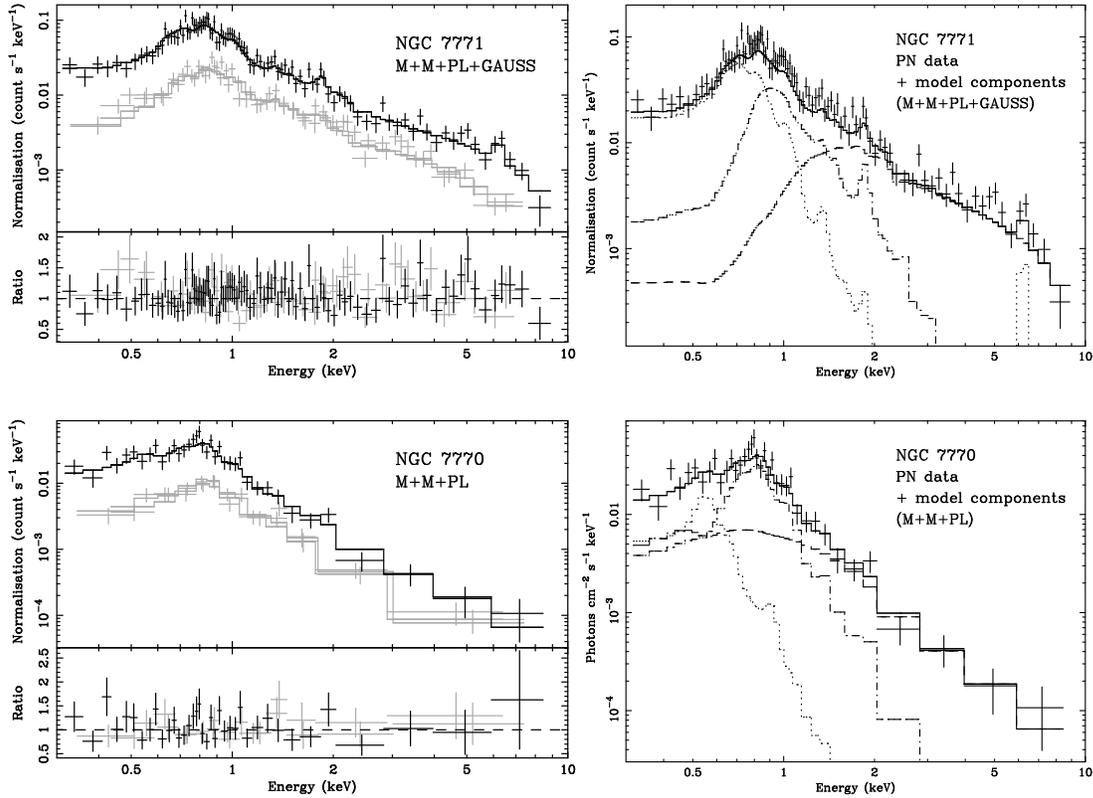

\centering
\rotatebox{270}{\includegraphics[width=5cm]{figure2a.ps}}
\vspace*{0.5cm}
\rotatebox{270}{\includegraphics[width=5cm]{figure2b.ps}}
\rotatebox{270}{\includegraphics[width=5cm]{figure2c.ps}}
\vspace*{0.5cm}
\rotatebox{270}{\includegraphics[width=5cm]{figure2d.ps}}
\caption{Top left: \xmm spectra and {\small M}+{\small M}+{\small PL}+GAUSS model for NGC~7771. PN data points and model fit are shown in black; those of the MOS data are shown in grey. Top right: The unfolded {\small M}+{\small M}+{\small PL}+GAUSS model and PN spectrum to illustrate the contributions from the model components. Bottom left: \xmm spectra and {\small M}+{\small M}+{\small PL} model for NGC~7770 and bottom right: the unfolded PN spectrum. The model components are denoted by dashed (power-law), dotted (cool {\small MEKAL}) \& dot--dashed (warm {\small MEKAL}) lines.}
\label{fig:spec7771}
\end{figure*}

We began the X-ray analysis of the central source in NGC~7771 by fitting the spectra with single- and two-component spectral models with combinations of power-law and {\small MEKAL} components, each with separate absorbing columns. All single-component models were rejected due to unacceptably high values of chi-squared ($\chi^2_{\nu}>2$), but good fits were obtained with two-component models. The soft component can be represented by a warm thermal plasma ($kT\sim0.5$\,keV), and the hard component is well modelled by either a hard power-law ($\Gamma\sim1.5$) or a hot thermal plasma ($kT\sim19$\,keV). Although these fits are statistically indistinguishable with $\chi^2_{\nu}\sim0.98$ in all cases, the most realistic interpretation of the data is the {\small M}+{\small PL} model, as we know that the hard 2--10\,keV emission in nearby SBGs is dominated by a small number of bright discrete sources which are likely to be XRBs. However, even though hotter gas (typically a few keV) from SNRs is also detectable in data of sufficient spatial and spectral quality for point source subtraction (e.g. NGC~3256, \citealt{lira02}), we discount the {\small M}+{\small M} models as the hot plasma temperatures ($kT\sim19$\,keV) are unrealistically high. Table~\ref{table:n7771} shows the results of the {\small M}+{\small PL} fits. We have included models where both absorbing components were were tied together (model 1) and where they were fitted independently (model 2), although there is little difference between them. 

However, although this simple model does adequately describe the data, we have also tried fitting the spectra with a more complex three component model ({\small M}+{\small M}+{\small PL}), as it is unrealistic to assume that the thermal plasma will possess one distinct temperature. Indeed, these systems are expected to possess thermal plasmas with a range of temperatures spanning $T\sim10^5-10^8$\,K \citep{stricklandstevens00}. This three-component model yields plasma temperatures of $kT$=0.4/0.7\,keV and a slightly softer power-law with $\Gamma$=1.7, with a 98.7 percent improvement over the {\small M}+{\small PL} model. We therefore adopt this as our best-fit model (model 3), and Figure~\ref{fig:spec7771} shows the EPIC data plus residuals (top left) and the additive model components with the PN data (top right). This model implies that, while there is no additional absorption associated with the cool {\small MEKAL} component above the Galactic value (4.35$\times10^{20}$ cm$^{-2}$, \citealt{dickey90}), the power-law continuum and warm thermal plasma component are much more heavily absorbed than in the simple {\small M}+{\small PL} models, although there are large uncertainties on these absorption parameters. Note that we were unable to improve the fit any further with the addition of more {\small MEKAL} components.

Table~\ref{table:n7771_comps} shows the relative contributions of the power-law and {\small MEKAL} components to the total X-ray emission in NGC~7771. The emission is dominated by the power-law component, indicating the presence of multiple XRBs or a low-luminosity AGN (LLAGN). The spectra for this source do show evidence of possible line emission at 6--7\,keV, which could be evidence for AGN activity (neutral Fe-K at 6.4\,keV) or emission from type Ib/IIa SNRs (Fe 6.7\,keV, e.g. \citealt{behar01}) in the central starburst region. To investigate this, we have attempted to fit a narrow gaussian line to the power-law continuum of the best-fit {\small M}+{\small M}+{\small PL} model. If all model parameters were left free, we were unable to constrain the line energy. However, restricting the line energy to fit between 6 and 7\,keV and freezing all other model components resulted in a line energy of 6.24$^{+0.28}_{-0.10}$\,keV with an equivalent width (EW) of 311\,eV, with a 90 per cent upper limit of 857\,eV  (plotted in Figure~\ref{fig:spec7771}, top). If the line energy was frozen at 6.4\,keV (i.e. at the expected position of a neutral Fe-K line from an AGN), the measured EW was 323\,eV with an upper limit of 915\,eV. The addition of this line only improves the fit statistics by $\sim1.7\sigma$; however its implications to the possibility of the presence of an LLAGN in NGC~7771 are discussed in section~\ref{sec:disc_llagn}.

\begin{table}
\caption{Fluxes, luminosities and percentage of total emission from the best-fit models of the central source in NGC~7771 and NGC~7770.}
 \centering
  \begin{tabular*}{8.4cm}{@{}lcccccccc@{}}
\hline
\\[-3mm]
Component  && $F_X$ && Per cent && $L_X$ && Per cent \\
           &&       && of total &&       && of total \\      
\\[-3mm]
\hline
\\[-3mm]
\multicolumn{9}{c}{\bf NGC~7771 Central Source {\small M}+{\small M}+{\small PL}}\\
\\[-3mm]
\hline
\\[-3mm]
Cool {\small MEKAL} && 0.44  && 16       && 0.18  && 8    \\
Warm {\small MEKAL} && 0.41  && 15       && 0.87  && 39   \\
Power-law  && 1.93  && 69       && 1.19  && 53   \\

\\[-2mm]
Total      && 2.78  && 100      && 2.24  && 100  \\
\\[-3mm]
\hline
\\[-3mm]
\multicolumn{9}{c}{\bf NGC~7770 {\small M}+{\small M}+{\small PL}}\\
\\[-3mm]
\hline
\\[-3mm]
Cool {\small MEKAL} && 0.06  && 10     && 0.10  && 24   \\
Warm {\small MEKAL} && 0.22  && 37     && 0.14  && 33   \\
Power-law           && 0.32  && 53     && 0.18  && 43   \\

\\[-2mm]
Total               && 0.60  && 100    && 0.42  && 100  \\
\\[-3mm]
\hline
\end{tabular*}
\begin{tabular*}{8.4cm}{@{}l}
Notes:  {\it Observed} fluxes in the 0.3--10\,keV band, in units of $10^{-13}$ \\erg s$^{-1}$ cm$^{-2}$. {\it Unabsorbed} luminosities in the 0.3--10\,keV band, in \\units of $10^{41} \ergsec$ (assuming a distance of 56.7\,Mpc).\\
\end{tabular*}
\label{table:n7771_comps}
\end{table}

\subsubsection{NGC~7770 Spectral Analysis}
\label{sec:n7770_spec}

The X-ray spectra of the dwarf companion NGC~7770 were fitted with the same set of spectral models, and the results are also shown in Table~\ref{table:n7771}. These data are well-fit with a {\small M}+{\small PL} model, although a statistically better fit is obtained by fitting the two absorbing columns separately (model 2). It is, however, unrealistic to have negligible absorption for the power-law component and substantial absorption for the thermal component as this model implies, since we assume that they originate from the same (star-forming) areas of the galaxy (i.e. the central and disc regions). Given the more limited statistics of the data, we therefore assume that both components have the same absorbing column (model 1, $kT$=0.5\,keV, $\Gamma$=2.4).

However, as in the case of the central source in NGC~7771, the {\small M}+{\small PL} fit can be improved with the addition of a second {\small MEKAL} component. Again, given the more limited statistics for this galaxy, a model with separate absorbing columns for each component yields unrealistic values of $N_H$, and so we have applied one column to all the components (model 3). Although partially unconstrained, we are able to fit a low-temperature thermal component with $kT$=0.15\,keV to the data, which improves the fit statistics at the 98.9 per cent (2.5$\sigma$) level.  The data and model are plotted in Figure~\ref{table:n7771}, and Table~\ref{table:n7771_comps} shows the contributions of the power-law and {\small MEKAL} components to the total X-ray emission for the {\small M}+{\small M}+{\small PL} model. In contrast to the central source in NGC~7771, the thermal components contribute (in total) a larger fraction of the unabsorbed luminosity than the power-law component. This agrees with the soft nature of the galaxy in the \xmm EPIC images (Figure~\ref{fig:n7771}). Since this galaxy does have such a soft spectrum, we have attempted to fit it with thermal components only. A two-temperature plasma model ({\small M}+{\small M}) gives an unsatisfactory fit to the data ($kT$=0.2/3.1\,keV, $\chi^2_{\nu}\sim1.4$). If however we replace the soft power-law in model 3 with a {\small MEKAL} component to fit the high energy end of the spectrum (i.e. {\small M}+{\small M}+{\small M}), this does yield a good fit ($\chi^2_{\nu}\sim0.9$) with plasma temperatures of $kT=0.16/0.56/4.00$\,keV. However this interpretation is problematic as the absence of a non-thermal component implies that there are no bright XRB systems present in NGC~7770.

\begin{table*}
\caption{Spectral fitting results for the two ULXs in NGC~7771.}
 \centering
  \begin{tabular}{lccccccccccc}        
\hline
$N_H$                  & $kT/T_{in}$            &  $\Gamma$              &  $\chi^2$/dof      & $\Delta\chi^2$$^a$  & 1-P(F-test)$^b$       & $F_X$$^c$          & \multicolumn{2}{c}{$L_X$$^d$} & & \multicolumn{2}{c}{Flux Frac$^e$} \\

\\[-3mm]
                       & ($keV$)                &                        &                    &                     & (\%)                   &                    &  Obs            & Unabs       & & Soft        & Hard  \\
\hline
\multicolumn{12}{c}{\bf NGC 7771 X-1}\\
\hline
\multicolumn{12}{l}{{\bf Model 1:} {\small PL} {\small ({\small WABS}*PO)}}\\
1.03$^{+0.57}_{-0.65}$ & -              & 1.72$^{+0.22}_{-0.24}$            & 26.6/16            & -                   & -                     & 5.11$^{+0.72}_{-1.22}$   & 1.97$^{+0.28}_{-0.47}$ & 2.25$^{+0.31}_{-0.65}$ & & -           & - \\
\\[-2mm]
\multicolumn{12}{l}{{\bf Model 2: {\small PL}+{\small M} {\small ({\small WABS}*(PO+MEKAL))}}}\\

{\bf 2.36$^{+3.43}_{-1.87}$} & {\bf 0.28$^{+0.28}_{-0.11}$} & {\bf 1.56$^{+0.44}_{-0.35}$} & {\bf 14.2/14} & {\bf 12.4} & {\bf 98.8}                  & {\bf 5.51$^{+0.60}_{-2.51}$} & {\bf 2.12$^{+0.23}_{-0.97}$} & {\bf 2.95$^{+0.25}_{-2.23}$} & & {\bf 0.10} & {\bf 0.90} \\

\hline
\multicolumn{12}{c}{\bf NGC 7771 X-2}\\
\hline
\multicolumn{12}{l}{{\bf Model 1:} {\small PL} {\small ({\small WABS}*PO)}}\\

$<$1.21                   & -                      & 1.89$^{+0.21}_{-0.25}$ & 21.7/13            & -                   & -                     & 3.12$^{+0.66}_{-1.22}$ & 1.20$^{+0.25}_{-0.47}$ & 1.28$^{+0.20}_{-0.47}$  & & -           & -   \\

\\[-2mm]
\multicolumn{12}{l}{{\bf Model 2: {\small PL}+{\small DISKBB} {\small ({\small WABS}*(PO+DISKBB))}}}\\

{\bf 3.18$^{+4.49}_{-2.51}$}  & {\bf 0.16$^{+0.12}_{-0.03}$} & {\bf 1.67$^{+0.27}_{-0.26}$} & {\bf 15.9/11} & {\bf 5.8}  & {\bf 82.0}  & {\bf 3.30$^{+0.45}_{-2.57}$} & {\bf 1.27$^{+0.17}_{-0.99}$} & {\bf 3.86$^{+0.10}_{-3.21}$} & & {\bf 0.20} & {\bf 0.80} \\

\\[-2mm]
\multicolumn{12}{l}{{\bf Model 3:} {\small PL}+{\small BBODY} {\small ({\small WABS}*(PO+BBODY))}}\\

3.23$^{+4.04}_{-1.61}$ & 0.12$^{+0.04}_{-0.04}$ & 1.73$^{+0.31}_{-0.54}$ & 16.0/11               & 5.7                 & 81.1                  & 3.26$^{+0.47}_{-2.70}$ & 1.25$^{+0.18}_{-1.04}$ & 3.50$^{+0.11}_{-3.17}$ & & 0.18 & 0.82 \\

\\[-2mm]
\multicolumn{12}{l}{{\bf Model 4:} {\small PL}+{\small M} {\small ({\small WABS}*(PO+MEKAL))}}\\

$<$1.98                   & 0.58$^{+0.52}_{-0.30}$ & 1.64$^{+0.20}_{-0.29}$ & 16.7/11            & 5                   & 76.1                  & 3.34$^{+0.69}_{-1.30}$ & 1.29$^{+0.26}_{-0.50}$ & 1.33$^{+0.23}_{-0.59}$ & & 0.08        & 0.92\\

\hline
\end{tabular}
\begin{tabular}{l}
Notes: ~Spectral models and parameters as in Table~\ref{table:n7771} except: {\small BBODY}=blackbody model and {\small DISKBB}=multicolour disc black body \\model. ~$^a$Improvement in the $\chi^2$ statistic over the single component model power-law ({\small PL}) fit, for two extra degrees of freedom. \\~$^b$F-test statistical probability of improvement of fit over single component model ({\small PL}). ~$^c${\it Observed} fluxes in the 0.3--10\,keV band, in \\units of $10^{-14}$ erg s$^{-1}$ cm$^{-2}$. ~$^d${\it Observed} and {\it unabsorbed} luminosities in the 0.3--10\,keV band, in units of $10^{40} \ergsec$ (assuming a \\distance of 56.7\,Mpc). ~$^e$Fraction of total flux in the hard ({\small PL}) and soft model components over 0.3--10\,keV band. The best-fit model \\for each source is highlighted in bold.\\
\end{tabular}
\label{table:ulx}
\end{table*}

\begin{figure*}
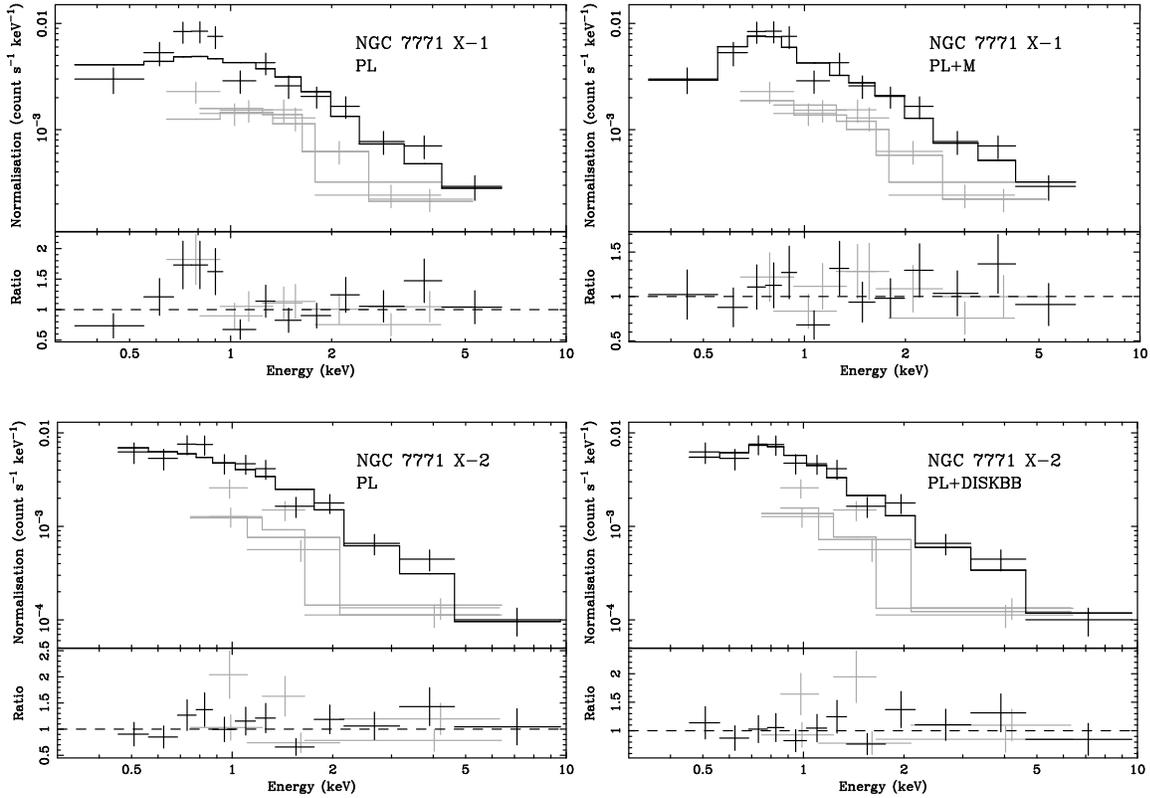

\centering
\rotatebox{270}{\includegraphics[width=5cm]{figure3a.ps}}
\vspace*{0.5cm}
\rotatebox{270}{\includegraphics[width=5cm]{figure3b.ps}}
\rotatebox{270}{\includegraphics[width=5cm]{figure3c.ps}}
\vspace*{0.5cm}
\rotatebox{270}{\includegraphics[width=5cm]{figure3d.ps}}
\caption{\xmm spectra of the two ULXs in NGC~7771. Top: {\small PL} (left) and {\small PL}+{\small M} (right) models for X-1. Bottom: {\small PL} (left) and {\small PL}+{\small DISKBB} (right) models for X-2. PN data points and model fit are shown in black; those of the MOS data are shown in grey.}
\label{fig:speculx}
\end{figure*}

\subsubsection{ULX Properties}
\label{sec:n7771_ulx}

The two bright discrete sources located at either end of the galactic bar in NGC~7771 (X-1 at $\alpha$=$23^h51^m22.3^s$, $\delta$=$+20\deg06\arcm37\arcs$; X-2 at $\alpha$=$23^h51^m27.2^s$, $\delta$=$+20\deg06\arcm52\arcs$) both have X-ray luminosities in the ULX regime ($>10^{39} \ergsec$), and both have sufficient counts for simple spectral fitting ($>300$ in the combined MOS \& PN data). 

To begin with, we fitted the spectra of both sources with simple absorbed single-component spectral models: {\small PL}, {\small MEKAL}, {\small DISKBB} and {\small BBODY}. Of these, only the absorbed {\small PL} models gave statistically acceptable fits with $\chi^2_{\nu}<2$ for both sources (Table~\ref{table:ulx}). However, since these fits still gave a $\chi^2_{\nu}$ well above 1, and both showed evidence for a soft excess (Figure~\ref{fig:speculx}, left), we attempted to improve the fits with the addition of a soft component ({\small MEKAL}/{\small DISKBB}/{\small BBODY}), the results of which are also shown in Table~\ref{table:ulx}. In the case of X-1, the best fit was achieved with a statistical improvement at the 98.8 percent (2.5$\sigma$) level with the addition of a soft {\small MEKAL} thermal component with $kT\sim0.3$\,keV and the resulting fit is shown in Figure~\ref{fig:speculx} (top right). Even though the addition of either {\small DISKBB} \& {\small BBODY} components also improved the fit, the resulting temperatures were very low ($\sim0.07$\,keV), and both models had large absorption columns which resulted in unrealistically high unabsorbed luminosities ($L_X>10^{42} \ergsec$). For X-2, the {\small PL} fit was only improved with the addition of a soft component at the $\sim1.3\sigma$ level (marginally best-fit with a {\small PL}+{\small DISKBB} model, Figure~\ref{fig:speculx}, bottom right), although this could simply be a consequence of the more limited photon statistics in this fainter source. 

\begin{figure}
\centering
\rotatebox{270}{\scalebox{0.6}{\includegraphics{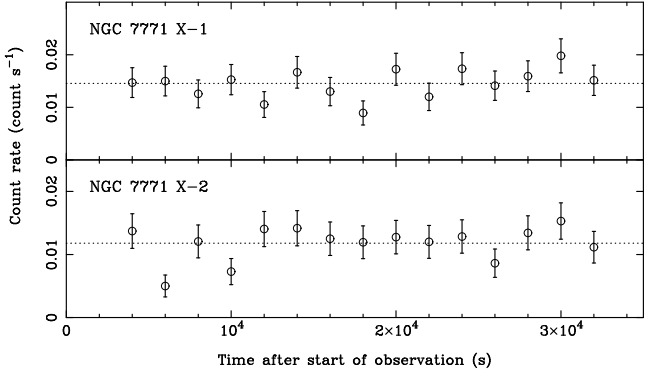}}}
\caption{\xmmn light curves for the two ULXs in NGC~7771, obtained by summing the counts from the 3 EPIC cameras. The mean count rate is shown as a dashed line and error bars correspond to 1$\sigma$ deviations assuming Gaussian statistics.}
\label{fig:lcurves}
\end{figure}

Each source was tested for variability over the duration of the \xmmn observation using both $\chi^2$ and Kolmogorov-Smirnov (K-S) tests. Firstly, the $\chi^2$ test was used to search for large amplitude variability in binned data against the hypothesis of a constant count rate, and for this purpose we derived short-term light curves for each source (Figure~\ref{fig:lcurves}). The data from the 3 EPIC cameras were co-added for improved signal-to-noise ratio, and tailored so that each bin had at least 20 counts after background subtraction, resulting in temporal resolutions of 2000s for each source. With this test, X-2 showed some sign of short-term variability with $\chi^2_{\nu}\sim1.9$, equivalent to a 97 percent ($\sim2.2\sigma$) probability of variability, while X-1 showed no evidence of variability ($\chi^2_{\nu}\sim1.1$). Secondly, the K-S test was used to search for smaller gradual small amplitude variations by comparing the observed background subtracted cumulative photon arrival distribution with the expected distribution if the flux was constant.  These tests were conducted using background subtracted PN light curves with a time-resolution of 1\,s. Here, it was X-1 that showed evidence of variability at the 97 percent ($\sim2.2\sigma$) level, while the data for X-2 showed none ($<1\sigma$).

\subsubsection{Comparison with previous observations}
\label{sec:n7771_comp}

NGC~7771/0 has previously been observed in X-rays using both \einstein and \rosat (\citealt{fabbiano92}; \citealt{rephaeli95}; \citealt{davies97}). \citet{davies97} estimated a total (0.16--3.5\,keV) luminosity of $L_X\sim2.3\times10^{41} \ergsec$ for the combined emission of NGC~7771 and NGC~7770 from the \einstein data (corrected for the distance used in this paper), assuming a Raymond-Smith (RS) thermal plasma model with $kT$=1.7\,keV. In the same study, \rosat PSPC images showed strong X-ray emission from the galaxy's centre, plus extended emission throughout the disc. Spectral analysis of the PSPC data was restricted due to the low number of counts ($\sim$100), and \citet{davies97} were unable to distinguish between power-law and RS thermal plasma models. The power-law model gave an X-ray luminosity of $6.8\times10^{40} \ergsec$ (0.1--2.4\,keV) for the combined emission of NGC~7771 and NGC~7770, which is consistent with the total observed luminosity of the system in the same band in the \xmmn observation ($L_X\sim7.6\times10^{40} \ergsec$).

\subsection{NGC~2342/2341}
\label{sec:n2342_results}

\subsubsection{X-ray morphology}
\label{sec:n2342_morph}

\begin{figure}
\centering
\scalebox{0.7}{\includegraphics{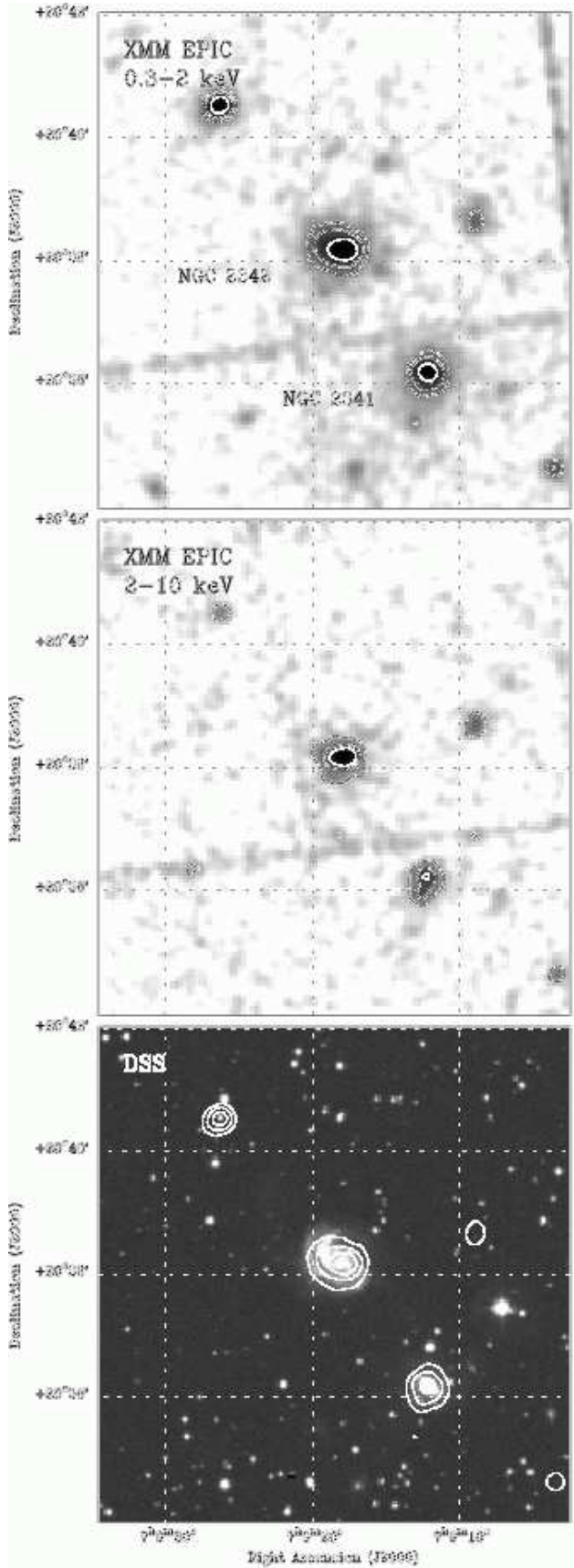}}
\caption{\xmmn EPIC soft (0.3--2\,keV, top) \& hard (2--10\,keV, middle) X-ray images of NGC~2342/1 with surface brightness contours overlaid. Bottom: DSS optical blue image of NGC~2342/1 with broad-band (0.3--10\,keV) X-ray contours overlaid. See text for further details.}
\label{fig:n2342}
\end{figure}

In Figure~\ref{fig:n2342} we show soft (0.3--2\,keV, top) and hard (2--10\,keV, middle) \xmmn EPIC images of the NGC~2342/1 pair of galaxies. The images are the sum of all 3 detectors, and are convolved with a 1$\sigma$ (4 arcsecond) 2-D Gaussian mask and displayed with a logarithmic scale. The intensity minima/maxima correspond to surface brightnesses of $8.5\times10^{-6}$/$5.4\times10^{-4}$ ct s$^{-1}$ (soft) and $8.5\times10^{-6}$/$2.7\times10^{-4}$ ct s$^{-1}$ (hard). The contours correspond to 0.7,1.4 \& $3.4\times10^{-4}$ ct s$^{-1}$ (soft) and 0.3,0.7 \& $1.3\times10^{-4}$ ct s$^{-1}$ (hard). We do not resolve any detailed structure in either galaxy with {\it XMM-Newton}, but the companion galaxy NGC~2341 does show an interesting elongated morphology in the hard X-ray image. Figure~\ref{fig:n2342} (bottom) shows a DSS blue optical image of the system overlaid with broad-band (0.3--10\,keV) X-ray contours corresponding to surface brightnesses of 0.9,2.0 \& 4.7$\times10^{-4}$ ct s$^{-1}$.

\subsubsection{NGC~2342 Spectral Analysis}
\label{sec:n2342_spec}

The integrated X-ray spectra for each of the galaxies were fitted using the same method described in section~\ref{sec:n7771_results}, and the results are shown in Table~\ref{table:n2342}. Although the spectra of NGC~2342 can be well-fit with a {\small M}+{\small PL} model, the addition of another {\small MEKAL} component improves the fit at the 98.3 percent (2.4$\sigma$) level (model 3). We have had to use a single absorbing column for all components, as applying separate columns to each component resulted in unrealistic values of $N_H$ (as with NGC~7770, see section~\ref{sec:n7770_spec}). This {\small M}+{\small M}+{\small PL} model (plotted in Figure~\ref{fig:spec2342}, top) yields  $kT\sim0.3/0.7$\,keV, $\Gamma\sim2.0$, and an absorption column of approximately twice the Galactic value (8.93$\times10^{20}$ cm$^{-2}$, \citealt{dickey90}) resulting in an unabsorbed luminosity of $L_X\sim2.7\times10^{41} \ergsec$. Note that in this case, the two-temperature thermal plasma model ({\small M}+{\small M}) does not give a satisfactory fit to the data ($\chi^2_{\nu}\sim1.5$). The relative contributions of the spectral components for the best-fit model are shown in Table~\ref{table:n2342_comps}; the power-law component is very dominant in this case.

\subsubsection{NGC~2341 Spectral Analysis}
\label{sec:n2341_spec}

The spectra of this galaxy are best-fit with a simple {\small M}+{\small PL} model; in this case, the addition of another {\small MEKAL} component did not improve the fit above the 70 per cent level. As with NGC~7770, even though a superior fit is obtained with a {\small M}+{\small PL} with separate absorbing $N_H$ columns for the two components (Table~\ref{table:n2342}, model 2), the statistics for this source are too limited to separate the absorption columns in this way. Again, it is unrealistic to have a cool thermal plasma component which is more absorbed than the power-law component, and we therefore adopt the {\small M}+{\small PL} model with a single absorbing column as the best-fit model (model 1, plotted in Figure~\ref{fig:spec2342}, bottom). The data are also well-fit ($\chi^2_{\nu}\sim1.1$) with the two-temperature thermal plasma ({\small M}+{\small M}) model with temperatures of $kT\sim0.5/7.8$\,keV (not shown in Table~\ref{table:n2342}), but we favour the power-law for the hard component, interpreted as emission from a population of XRBs. In this case, the contributions to the unabsorbed luminosity from the two components in the best-fit model are comparable (Table~\ref{table:n2342_comps}), with an enhanced percentage from the {\small MEKAL} component compared to NGC~2342.

\subsubsection{Comparison with previous observations}
\label{sec:n2342_comp}

The only other X-ray observation of this system was performed with the \rosat PSPC \citep{henriksen99}. However, the spectrum was extracted using a circular region of 3.4 arcminutes radius, which included the emission from both galaxies. These data were fit with RS thermal plasma models in the 0.25--2\,keV range, giving a best-fit temperature of $kT\sim1.1$\,keV and a luminosity of $L_X\sim1.9\times10^{41} \ergsec$ (corrected for the distance used in this paper). The total X-ray luminosity of both galaxies in the \xmm data in the same energy range is $L_X\sim1.2\times10^{41} \ergsec$; this $\sim$ 35 percent reduction is likely to be due to the large extraction region used in the analysis of the PSPC data, rather than a genuine change in the X-ray output of the galaxies themselves, although the latter possibility cannot be excluded.

\begin{table*}
\caption{Spectral fitting results for NGC~2342/2341.}
 \centering
  \begin{tabular}{lccccccccccc}
\hline
\multicolumn{2}{c}{{\small PL}} & & \multicolumn{2}{c}{M$_1$} & & \multicolumn{2}{c}{M$_2$}   &  $\chi^2$/dof  & $F_X$$^b$     & \multicolumn{2}{c}{$L_X$$^c$}   \\ 

\\[-3mm]

$N_H$$^a$              & $\Gamma$  & & $N_H$$^a$ & $kT (keV)$    & & $N_H$$^a$  & $kT (keV)$  & &              &  Obs            & Unabs         \\
\hline

\multicolumn{12}{c}{\bf NGC~2342}\\
\hline
\multicolumn{12}{l}{{\bf Model 1:} {\small M}+{\small PL} {\small (WABS*(PO+MEKAL))}}\\

1.98$^{+0.21}_{-0.36}$ & 2.07$^{+0.11}_{-0.12}$ & & $\dag$ & 0.33$^{+0.06}_{-0.05}$ & & - & - &144.7/158 & 2.94$^{+0.22}_{-0.28}$ & 1.77$^{+0.13}_{-0.17}$ & 2.82$^{+0.13}_{-0.52}$ \\

\\[-2mm]

\multicolumn{12}{l}{{\bf Model 2:} {\small M}+{\small PL} {\small (WABS*PO+WABS*MEKAL)}}\\

1.70$^{+0.34}_{-0.34}$ & 2.02$^{+0.12}_{-0.12}$ & & $<$1.46                  & 0.62$^{+0.06}_{-0.15}$ & & - & - & 144.1/157 & 2.96$^{+0.10}_{-0.33}$ & 1.79$^{+0.06}_{-0.20}$ & 2.47$^{+0.03}_{-0.46}$\\

\\[-2mm]

\multicolumn{12}{l}{{\bf Model 3: {\small M}+{\small M}+{\small PL} {\small (WABS*(PO+MEKAL+MEKAL))}}}\\

{\bf 1.85$^{+0.41}_{-0.36}$} & {\bf 1.99$^{+0.13}_{-0.10}$} & & $\dag$ & {\bf 0.29$^{+0.09}_{-0.05}$} & & $\dag$ & {\bf 0.74$^{+0.19}_{-0.13}$}  & {\bf 137.3/156} & {\bf 2.98$^{+0.23}_{-0.28}$} & {\bf 1.80$^{+0.14}_{-0.17}$} & {\bf 2.73$^{+0.17}_{-0.40}$} \\

\hline
\multicolumn{12}{c}{\bf NGC~2341}\\
\hline

\multicolumn{12}{l}{{\bf Model 1: {\small M}+{\small PL} {\small (WABS*(PO+MEKAL))}}}\\

{\bf 2.83$^{+1.66}_{-1.45}$} & {\bf 1.82$^{+0.14}_{-0.09}$} & & $\dag$ & {\bf 0.30$^{+0.09}_{-0.06}$} & & - & - & {\bf 92.3/87} & {\bf 1.51$^{+0.15}_{-0.30}$} & {\bf 0.91$^{+0.09}_{-0.18}$} & {\bf 1.79$^{+0.18}_{-0.85}$} \\

\\[-2mm]
\multicolumn{12}{l}{{\bf Model 2:} {\small M}+{\small PL} {\small (WABS*PO+WABS*MEKAL)}}\\

1.87$^{+1.71}_{-0.84}$ & 1.70$^{+0.14}_{-0.18}$ & & 3.62$^{+2.01}_{-1.60}$ & 0.29$^{+0.05}_{-0.07}$ & & - & - &90.6/86 & 1.54$^{+0.11}_{-0.35}$ & 0.93$^{+0.06}_{-0.21}$ & 2.01$^{+0.13}_{-1.05}$\\

\hline
\end{tabular}
\begin{tabular}{l}
Notes: ~Spectral models and parameters as in Table~\ref{table:n7771}. ~$^a${\it Observed} fluxes in the 0.3--10\,keV band, in units of $10^{-13}$ erg s$^{-1}$ cm$^{-2}$. \\~$^b${\it Observed} and {\it unabsorbed} luminosities in the 0.3--10\,keV band, in units of $10^{41} \ergsec$ (assuming a distance of 71\,Mpc). ~$\dag$Same \\hydrogen column as applied to the previous spectral component. Errors correspond to 90 per cent confidence limits for parameter of \\interest. The best-fit models are highlighted in bold.\\ 
\end{tabular}
\label{table:n2342}
\end{table*}

\begin{figure*}
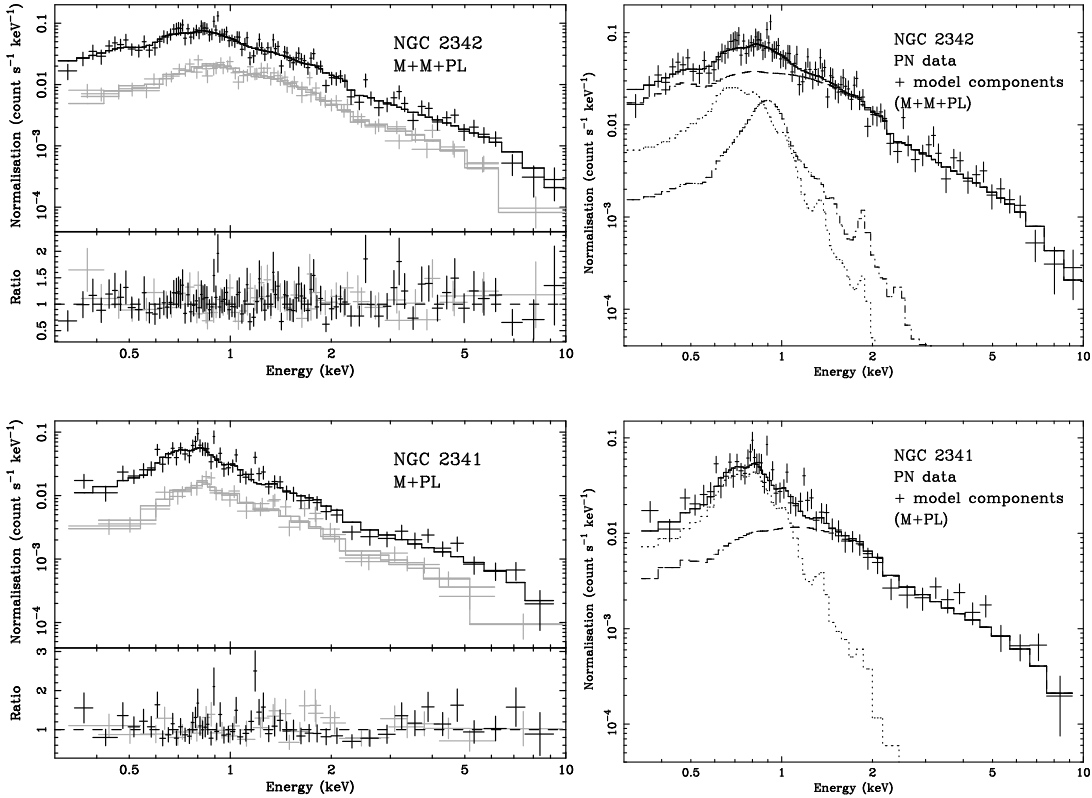

\centering
\rotatebox{270}{\includegraphics[width=5cm]{figure6a.ps}}
\vspace*{0.5cm}
\rotatebox{270}{\includegraphics[width=5cm]{figure6b.ps}}
\rotatebox{270}{\includegraphics[width=5cm]{figure6c.ps}}
\vspace*{0.5cm}
\rotatebox{270}{\includegraphics[width=5cm]{figure6d.ps}}
\caption{Top left: \xmm spectra and {\small M}+{\small M}+{\small PL} model for NGC~2342. PN data points and model fit are shown in black; those of the MOS data are shown in grey. Top right: The unfolded model and PN spectrum to illustrate the contributions from the model components. Bottom left: \xmm spectra and {\small M}+{\small PL} model for NGC~2341, and bottom right: the unfolded PN spectrum. The model components are denoted by dashed (power-law), dotted (cool {\small MEKAL}) \& dot--dashed (warm {\small MEKAL}) lines.} 
\label{fig:spec2342}
\end{figure*}

\begin{table}
\caption{Fluxes, luminosities and percentage of total emission from the best-fit models of NGC~2342 and NGC~2341.}
 \centering
  \begin{tabular*}{8.4cm}{@{}lcccccccc@{}}
\hline
\\[-3mm]
Component  && $F_X$ && Per cent && $L_X$ && Per cent \\
           &&       && of total &&       && of total \\
\\[-3mm] 
\hline
\\[-3mm]
\multicolumn{9}{c}{\bf NGC~2342 {\small M}+{\small M}+{\small PL}}\\
\\[-3mm]
\hline
\\[-3mm]
Cool {\small MEKAL} && 0.19  && 6       && 0.32  && 12  \\
Warm {\small MEKAL} && 0.15  && 5       && 0.16  && 6   \\
Power-law  && 2.64  && 89      && 2.25  && 82  \\

\\[-2mm]
Total      && 2.98  && 100     && 2.73  && 100 \\
\\[-3mm]
\hline
\\[-3mm]
\multicolumn{9}{c}{\bf NGC~2341 {\small M}+{\small PL}}\\
\\[-3mm]
\hline
\\[-3mm]
{\small MEKAL} && 0.34  && 23       && 0.79  && 44   \\
Power-law  && 1.17  && 77       && 1.00  && 56   \\

\\[-2mm]
Total      && 1.51  && 100      && 1.79  && 100  \\
\\[-3mm]
\hline
\end{tabular*}
\begin{tabular*}{8.4cm}{@{}l}
Notes:  {\it Observed} fluxes in the 0.3--10\,keV band, in units of $10^{-13}$ \\erg s$^{-1}$ cm$^{-2}$. {\it Unabsorbed} luminosities in the 0.3--10\,keV band, in \\units of $10^{41} \ergsec$ (assuming a distance of 71\,Mpc).\\
\end{tabular*}
\label{table:n2342_comps}
\end{table}

\section{Discussion}
\label{sec:discuss}

\subsection{The origin of the X-ray spectral components}

\citet{persic02} have recently used the spectral properties of M82 and NGC~253 as observed with {\it BeppoSAX} together with stellar-population evolutionary models to quantify the X-ray spectral components of SBGs. Contributions to the X-ray flux in the 0.3--10\,keV band are expected to come from XRBs, SNRs, Compton scattering of ambient FIR photons off supernova-accelerated relativistic electrons, diffuse thermal plasma and a compact nucleus in the form of a starburst or AGN. At low energies ($\la$2\,keV), the dominant component is expected to be diffuse thermal plasma resulting primarily from shock heating via the interaction of the low-density wind with the ambient high-density ISM \citep{strickland00a}. Above 2\,keV, they predict that the spectrum will be dominated by power-law emission from bright neutron star XRBs, plus a possible contribution from non-thermal Compton emission or an AGN if present \citep{persic02}. 

These predictions have in part been confirmed by \xmmn and {\it Chandra} observations of local SBGs, e.g. M82 (\citealt{matsumoto01}; \citealt{stevens03}), NGC~253 (\citealt{pietsch01}; \citealt{strickland00a}; \citealt{strickland02}), NGC~3256 \& NGC~3310 (\citealt{lira02}; \citealt{jenkins04b}) and NGC~4038/4039 (the Antennae; e.g. \citealt{fabbiano01}). Their broad band spectra can indeed be fit with single- or multi-temperature thermal plasmas with temperatures ranging between $\sim$0.2--0.9\,keV, plus a hard power-law tail at energies $>$2\,keV. The interpretation of the origin of the power-law component must however be revised; the high spatial resolution of \chandra has shown that the hard emission can be attributed to a few dominant luminous compact sources which are most likely XRBs with black-hole rather than neutron star primaries (based on their high luminosities). Spectacular examples of the detection of such sources include the Antennae \citep{zezas02}, the Cartwheel (\citealt{gao03}; \citealt{wolter04}) and Arp299 \citep{zezas03}.

Our spectral results for NGC~7771/0 and NGC~2342/1 are consistent with the other observations of nearby SBGs listed above. For both NGC~7771 and NGC~7770, we are able fit the thermal spectra with a two-temperature {\small MEKAL} models with $kT$=0.4/0.7\,keV and $kT$=0.2/0.5\,keV. In the case of NGC~7771, the trend of increasing absorption with increasing plasma temperature is similar to that found in other SBGs, and is evidence that the hotter component is likely to originate from supernova-heated gas in the most obscured central disc and starburst regions. A clear illustration of such a scenario is found in NGC~3256, where the hotter of the two plasmas can be spatially associated with the central $\sim$ 2\,kpc absorbed region of the galaxy, with the cooler component dominating in the outer regions most likely in the form of an out-flowing galactic wind \citep{lira02}. The complexity of the thermal models we are able to fit is highly dependent on the quality of the data, and we must bear in mind that these models are an over-simplification of the gas profiles in these galaxies i.e. it is physically unrealistic for there to be such distinct thermal plasma temperatures. This comparison of spectral fits of differing quality also highlights another point; with good enough statistics we can separate out the absorbing columns associated with each thermal component, whereas this type of modelling can lead to unrealistic values of $N_H$ when fit to spectra with insufficient counts (i.e. NGC~7770/2342/2341).

In addition to the soft thermal components, we can use the information gleaned from \chandra observations of other SBGs to interpret the origin of the hard ($\ga$ 2\,keV) integrated emission we detect from NGC~7770, NGC~2342 and NGC~2341, which are spatially unresolved by {\it XMM-Newton}. In these galaxies, there is no evidence for hidden luminous AGN activity, which can be detected at X-ray energies by the presence of an absorbed power-law ($N_H>10^{22} \cms$) and/or a strong neutral iron line at $\sim$6.4\,keV from the nucleus. In these cases, therefore, the power-law emission is almost certainly coming from populations of XRBs. However, there is some evidence for AGN activity in NGC~7771 (see section~\ref{sec:disc_llagn}).

\subsubsection{The ULXs in NGC~7771}
\label{sec:disc_ulx}

The two bright extra-nuclear X-ray sources in NGC~7771 are co-spatial with the ends of the galactic bar (see section~\ref{sec:n7771_morph}). This is consistent with models that show that gas can accumulate at the tips of bars due to the corotation of the bar structure with the disc \citep{englmaier97}, thereby creating star-forming regions where XRB sources can form.  A few other examples of X-ray sources at bar-ends have also been observed, for example in \rosat observations of the barred spirals NGC~1672 (\citealt*{brandt96}; \citealt{denaray00}) and {\it ROSAT/Chandra} observations of NGC~4303 (\citealt*{tschoke00}; \citealt{jimenez03}). 

The X-ray spectral fits to the ULXs are in the form of a power-law continuum plus a soft excess. X-1 is well fit with a {\small PL}+{\small M} model ($\Gamma\sim1.6$, $kT\sim0.3$\,keV, $\chi^2_{\nu}\sim1.0$), while the power-law fit to X-2 is marginally improved (although only at the 82 per cent level) with a {\small DISKBB} component ($\Gamma\sim1.7$, $T_{in}\sim0.2$\,keV, $\chi^2_{\nu}\sim1.4$), although normal {\small BBODY} and {\small MEKAL} components also give improved fits over a power-law only model. These types of spectral fits are commonly seen in bright ULXs (e.g. \citealt{miller03}; \citealt{jenkins04a}; \citealt{roberts04}). The physical nature of ULXs has been the topic of much debate in recent years, but \xmmn and \chandra observations are greatly improving our understanding of this phenomenon. The observed link between ULXs and episodes of high star formation, coupled with their spectral and timing properties suggest that the majority are likely to be high-mass XRBs going through a period of thermal-timescale mass transfer \citep{king04}. However, alternate explanations invoke, for example, the presence of intermediate-mass black holes (IMBHs, e.g. \citealt{colbert99}; \citealt{miller03}). Indeed, the possible presence of a such a soft accretion disc component in X-2 is consistent with the IMBH scenario, based on the argument that black hole mass scales inversely with the accretion disc temperature ($T_{in}\propto M^{-1/4}$, e.g. \citealt{makishima00}). However, the data are photon-limited leading to ambiguous results (i.e. that the soft emission can also be well-modelled with normal blackbody and thermal plasma models), and the {\small DISKBB} component is not is not significant enough for us to claim this as an IMBH candidate at present. The soft component in the spectrum of X-1, again at low spectral resolution, can be modelled by a thermal plasma which is likely to originate in the star-forming region where the source resides in a similar manner to the thermal plasma component in the central source in NGC~7771. The short-term variability detected in both sources (section~\ref{sec:n7771_ulx}) strongly suggests that their X-ray emission is dominated by single accreting XRBs, although high spatial resolution imaging with \chandra would be required to confirm this.

\subsection{Does NGC~7771 host a low-luminosity AGN?}
\label{sec:disc_llagn}

NGC~7771 possesses a large-scale bar and circumnuclear ring, both of which are likely to be a consequence of disc instabilities brought on by its interaction with NGC~7770 (\citealt{smith99}). The presence of a bar plays an important role in a galaxy's evolution; bars can extract angular momentum from the ISM in the disc through gravitation torques, causing gas to be driven into the inner few hundred pc of the galaxy, thereby fuelling a burst of star formation and possibly an AGN (e.g. \citealt{barnes91}). Observational evidence clearly shows that nuclear starburst activity is enhanced in galaxies with bars, often in the form of circumnuclear rings resulting from gas accumulating in the vicinity of inner Lindblad resonances (ILRs; \citealt*{knapen02}; \citealt{jogee04} and references therein). However, there is conflicting evidence with regards to the incidence of AGN in barred galaxies; while some studies find a higher fraction of bars in Seyfert galaxies compared with `normal' (non-active) galaxies (e.g. \citealt*{knapen00}; \citealt*{laurikainen04}), other results show that bars have negligible effect on the presence of an AGN (e.g. \citealt{mulchaey97}; \citealt*{ho97a}). This may be explained by the fact that models of gas flow in barred galaxies show that gas driven toward the nucleus can be interrupted by a nuclear ring, which may prevent the fueling of an AGN, although sufficient stellar material may be present in the immediate vicinity of galactic nuclei to fuel LLAGN (\citealt{ho97a} and references therein). However, since the duty cycles of AGN are expected to be a factor 10--100 times shorter than the lifetime of bars ($\sim$ 1\,Gyr), we do not expect all barred galaxies to show signs of AGN activity \citep{jogee04}. 

The lack of a correlation between bars and AGN may in part be explained by obscuration, i.e. `hidden' AGN. \citet*{maiolino99} reported that highly obscured AGN (e.g. Compton-thick Seyfert~2; N$_{H}>10^{24} \cms$) are found preferentially within strongly barred galaxies, whereas less obscured AGN are located in hosts with weak bars or no bar present. This correlation indicates that gas and dust driven towards the nucleus creates an obscuring screen which prevents us from viewing the nucleus directly.  In addition, starbursts are naturally expected to accompany the formation of an AGN due to the large amounts of gas in-flowing into the central regions, and it is feasible that the starburst itself (if extended) can obscure the AGN, resulting in optical spectra that appear to us as \hii-type objects \citep{fabian98}. Clear examples of this phenomena are NGC~4945 \citep{guainazzi00} and NGC~3690 \citep{dellaceca02}, both of which show evidence of AGN activity at X-ray wavelengths, although their optical and IR spectra are dominated by the starburst component and are classified as H{\small II}.

\citet{maiolino03} report that many hidden AGN are found in galaxies with $10^{11}<L_{IR}/L_{\odot}<10^{12}$, the range in which NGC~7771 lies. Even though no spectral signatures of AGN activity have been found at other wavelengths, a compact nuclear source is detected at radio and IR wavelengths \citep{smith99}. The presence in the new X-ray data of both a hard absorbed power-law component with a slope of $\Gamma\sim$ 1.7 typically seen in Seyfert galaxies \citep{nandra94}, plus a possible iron line at $\sim$ 6\,keV in the central source in NGC~7771 suggests that an obscured AGN may be present. We can test this hypothesis in several different ways. The absorption we measure in the X-ray data of $N_H\sim10^{22} \cms$ is equivalent to $A_V\sim5.5$ (assuming a standard Galactic value of E$_{{\rm B-V}}/N_H$; \citealt*{bohlin78}). The optical colour excess measured by \citep{veilleux95} implies an extinction consistent with this of $A_V\sim6.5$.  Using the relation of \citet*{elvis84} between the hard (2--10\,keV) X-ray luminosity and broad H$\alpha$ emission-line luminosity in LLAGN ($L_X/L(H\alpha)\simeq40$), we can determine whether we would expect to detect a broad H$\alpha$ line from the nucleus if one were present, or if it would be entirely obscured. The 2--10\,keV flux of NGC~7771 of $\sim1.8\times10^{-13} \ergcms$ predicts a broad H$\alpha$ component with $F(H\alpha)\sim4.5\times10^{-15} \ergcms$, which is a factor $\sim$ 6 less than the flux of the narrow line detected in the most obscured region of the nucleus by \citet{davies97}. With 6 magnitudes of extinction, this flux would be reduced by a factor $\sim$ 250, so a broad H$\alpha$ component would be beyond detection limits. There is also a UV source in the nucleus detected by IUE with a flux of $\sim2.5\times10^{-12} \ergcms$ (see figure~8, \citealt{davies97}). However, this emission is likely to be coming from the starburst ring ($\sim$ 6 arcsec diameter), where the obscuration is patchy \citep{smith99}, since the ratio of IR to UV flux on the spectral energy distribution (SED) of a factor of $\sim$ 5 implies a maximum A$_V$ of $\sim$ 1 magnitude in the UV if the intrinsic SED were that of an AGN. This is inconsistent with the extinction measured in the optical in the central $\sim$ 2 arcseconds of the galaxy. 

The final tentative piece of evidence for a LLAGN is the low-significance (1.7$\sigma$) detection of a narrow Fe-K$\alpha$ line at $\sim$6\,keV. We can predict the EW of such a line from an active nucleus using the empirical relation of \citet{page04}. This describes the X-ray `Baldwin effect', in which the EW of the line decreases with increasing luminosity (EW$\propto L_{(2-10 keV)}^{-0.17}$), possibly resulting from a decrease in the covering factor of the a molecular torus around the AGN. The 2--10\,keV luminosity of $L_X\sim7\times10^{40} \ergsec$ corresponds to an Fe-K$\alpha$ EW of $\sim$ 240\,eV, which is consistent with the measured (although partially unconstrained) EW of the line in the spectrum of the central source in NGC~7771. Another important point is that while the 90 per cent errors on the line energy (6.24$^{+0.28}_{-0.10}$\,keV) are consistent with 6.4\,keV emission, they do rule out the possibility that it originates from $\sim$ 6.7\,keV helium-like iron from the starburst component. Future X-ray observations with high signal-to-noise data at $\sim$ 6\,keV will be required to confirm the presence of this line, which could firmly establish the existence of a LLAGN in NGC~7771.

\subsection{The consequences of interaction}
\label{sec:interactions}

We know that interactions between galaxies increase their levels of star formation as seen in the optical/IR (see section~\ref{sec:intro}).  To investigate how the X-ray properties of our interacting systems relate to those of other galaxies, we have compared them with three different samples; isolated galaxies, interacting galaxies and merger systems. We note that these are in no sense complete samples, but are simply representative samples for comparison purposes. The X-ray data for the isolated galaxy sample (M101, NGC~278, M74, NGC~1291, NGC~2681, NGC~3184, NGC~4314, M94, M83 \& IC~5332) and the interacting system M51A/B are taken from the \chandra survey of nearby spiral galaxies of \citet{kilgard04}. The remainder are taken from {\it XMM-Newton}, \chandra or \rosat/\asca observations in the literature. In addition to the galaxies studied in this paper, the interacting systems are: The Cartwheel \citep{wolter04}; M82/M81/NGC~3077 (\citealt{stevens03}; \citealt{swartz03}; \citealt*{ott03}), NGC~4485/90 \citep{roberts02}, while the sample of merging galaxies comprises NGC~3690 (Arp~299; \citealt*{zezas98}), NGC~3256, NGC~3310 \& NGC~4438/39 (taken from \citealt{jenkins04b}), Arp~220 \citep{mcdowell03}, NGC~3921 \& NGC~7252 \citep{nolan04} and NGC~4676 \citep{read03}. For consistency between the datasets, these X-ray luminosities have all been converted to the 0.3--10\,keV band.

\begin{figure}
\centering
\rotatebox{270}{\scalebox{0.95}{\includegraphics{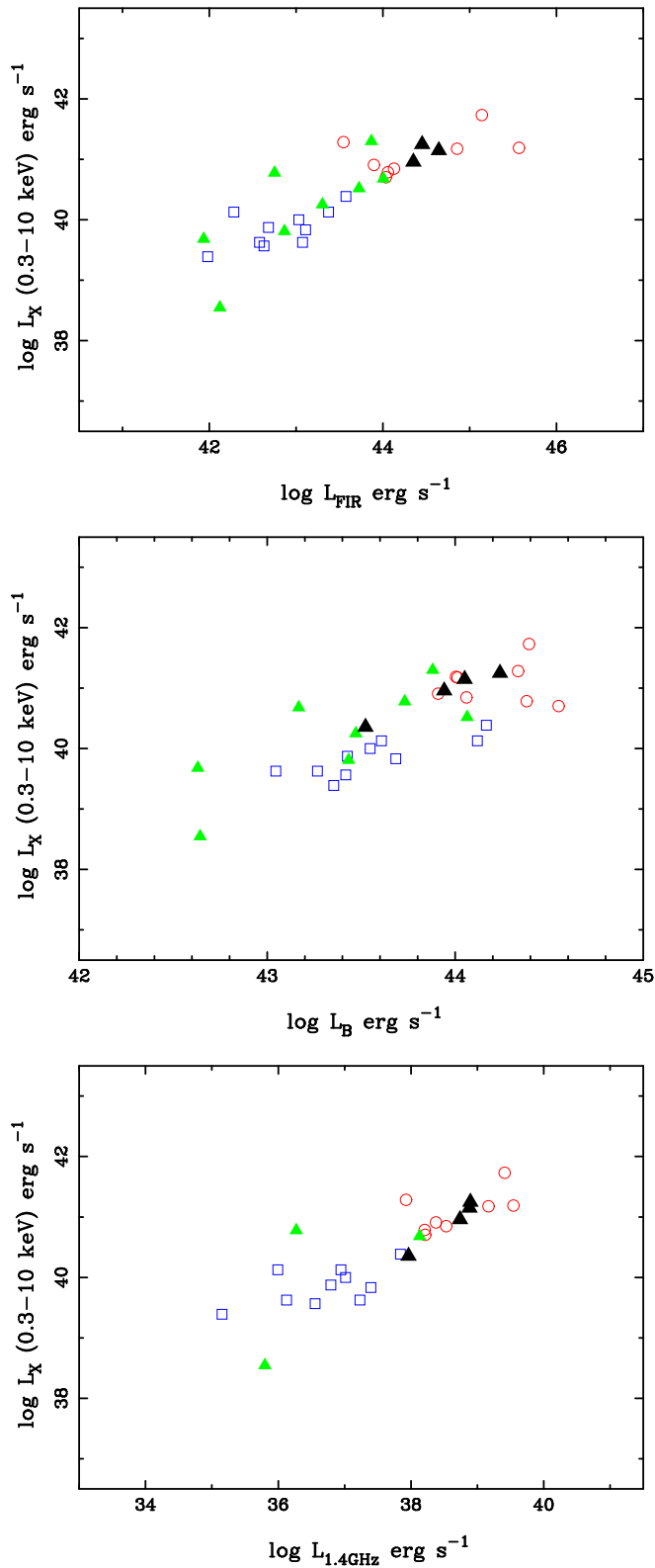}}}
\caption{Plots of the 0.3--10\,keV X-ray luminosity against FIR (top), blue (middle) and radio (bottom) luminosities for the three samples of galaxies. Isolated galaxies are marked with open squares, merging systems are marked with open circles, interacting galaxies are marked with filled triangles (with NGC~7771/0 and NGC~2342/1 marked with black triangles). See text for details of data sources. Note that NGC~7770 is not shown in the $L_X$/$L_{FIR}$ plot as there are no IR fluxes catalogued for this galaxy.}
\label{fig:lumplots}
\end{figure}

The FIR luminosities for each galaxy are calculated using FIR fluxes derived from the 60 and 100\,$\mu$m flux densities of \citet{sanders03}, \citet*{surace04} and the IRAS point source catalogue v2.0 (via the NED archive), using the relation of \citet*{helou85}:

\[
FIR=1.26\times10^{-11}(2.58S_{60_{\mu}} + S_{100_{\mu}}) \ergcms
\]

The optical B-band luminosities are calculated using the expression of \citep{tully88} relating the blue apparent magnitude $B_T$ and the distance $D$ (in Mpc):

\[
logL_B(L_{\odot})=12.192-0.4B_T+2logD
\]

The blue apparent magnitudes are mainly taken from the RC3 catalogue \citep{devaucouleurs91}, with the value for the Cartwheel taken from the ESO/Uppsala survey of the ESO(B) Atlas \citep{lauberts82}. The radio luminosities are derived from the 1.4\,GHz flux densities of \citet{condon98} where available, while the remainder are derived from the 1.49\,GHz flux densities from \citet{condon87} and \citet{condon90}.

The 0.3--10\,keV X-ray luminosities are plotted against these quantities in Figure~\ref{fig:lumplots} for the three galaxy samples. The majority of the sample show a strong correlation between the X-ray luminosity and their multiwavelength properties. This is to be expected, since the optical blue flux is consistent with the presence of young stellar populations, and both FIR and radio luminosities are strong tracers of the star formation activity associated with those populations. The FIR emission from SBGs comes primarily from dust heating by UV flux from massive OB stars in the starburst regions, which is re-radiated in the IR, while the nonthermal 1.4\,GHz radio flux is believed to arise from synchrotron radiation created by the acceleration of cosmic-ray electrons by supernova explosions (plus a small fraction from SNRs themselves) (e.g. \citealt{helou85}; \citealt*{condon91}).  X-ray emission is also an excellent tracer of star formation, since it arises from phenomena related to the end points of stellar evolution, mainly in the form of  XRBs, SNRs and diffuse thermal plasma.  A correlation between the FIR and $\sim$0.2--4.5\,keV X-ray emission in spiral and bright IR galaxies was first seen using \einstein (\citealt*{david92}; \citealt{fabbiano02}). More recently, \citet*{ranalli03} and \citet*{gilfanov04} have demonstrated that there is a tight linear correlation in SBGs between their FIR, radio and soft (0.5--2\,keV) and hard (2--10\,keV) X-ray luminosities, which we extend here to include isolated (non-starburst) galaxies. When the correlation between the FIR and 1.4\,GHz luminosities was reported by \citet{helou85}, it was also noted that it held over a wide range of objects, from quiescent spirals to galaxies dominated by nuclear starbursts. This agrees with extension of the isolated spiral galaxies in the X-ray correlations shown here.

These relations are continuous between the isolated, interacting and merging systems. However, while there is a clear segregation between the isolated and merging galaxies, the interacting sample span almost the entire range in luminosities at all wavelengths, even extending up to the level of some of the merging systems. The three interacting galaxies that appear to deviate from these correlations with X-ray luminosities roughly an order of magnitude higher than the rest of the sample in the $L_X$--$L_{FIR}$ plot are M81, whose X-ray luminosity is dominated by its bright nucleus \citep{swartz03}, the Cartwheel, which has numerous bright ULXs in a starburst ring away from the obscured centre of the galaxy \citep{wolter04}, and NGC~4485, whose X-ray output is dominated by one bright ULX \citep{roberts02}. 

\citet{ranalli03} find that while some galaxies with known LLAGN have $L_X$--$L_{FIR}$ and $L_X$--$L_{1.4GHz}$ ratios exceeding those of SBGs, some do follow the starburst relation, demonstrating that their outputs at these wavelengths are dominated by the star formation activity rather than the AGN component. An example of this is seen in NGC~3690 (Arp~299), where the emission from the buried AGN only contributes a very small fraction of the bolometric emission of the galaxy (\citealt{dellaceca02}; \citealt{zezas03}). It is clear from these plots that the galaxies studied in this paper (marked with black triangles) are among the brightest interacting galaxies in X-rays, and are at the high-luminosity end in all other properties, overlapping with some merging systems. In addition, these galaxies closely follow the SBG relation, demonstrating that their X-ray and FIR luminosities are dominated by the star formation activity, including NGC~7771 where there is a question as to the presence of a LLAGN.

\section{Summary \& Conclusions}
\label{sec:conclusions}

In this paper, we have studied the X-ray properties of two pairs of interacting galaxies, NGC~7771/0 and NGC~2342/1, in the 0.3--10\,keV energy band. For the first time, with the high quality spectra obtained with \xmmn we have detected the classic X-ray signatures of starburst activity in all four galaxies, in the form of soft multi-temperature thermal emission dominating below 2\,keV and hard power-law components representing the emission at higher energies (2--10\,keV).  

In NGC~7770, NGC~2342 and NGC~2341, we do not resolve any X-ray structure, but their integrated emission can be well-modelled with multi-temperature thermal plasmas with temperatures range between $kT\sim$ 0.2--0.7\,keV, with the warm component possibly representing supernova-heated gas within the absorbed discs and central regions and the cooler components possibly representing out-flowing material from the starburst interacting with the ISM. The power-law components have slopes of $\Gamma\sim$1.8--2.3, and are likely to represent the integrated emission from populations of XRBs associated with the starbursts, similar to those seen in other nearby merger systems.  
In NGC~7771, we resolve the X-ray emission into a bright central source plus two ULXs located at the either end of its bar at the corotation radius. The soft X-ray spectrum of the central source is well-modelled by a two-temperature thermal plasma with $kT\sim$0.4/0.7\,keV, showing increasing absorption with increasing plasma temperature. However, the power-law component is relatively flat and absorbed ($\Gamma\sim1.7$, $N_H\sim10^{22} \cms$). This, coupled with a low-significance detection of an emission line with an EW $\sim$300\,eV at $\sim$6\,keV, is the first indication that NGC~7771 may host a LLAGN, perhaps fuelled by the inflow of material along the galactic bar. The two bright bar sources are well-fit with power-law plus soft excess models commonly seen in ULXs, and both are variable above the 2$\sigma$ level, implying that their emission is dominated by single accreting XRBs.

A comparison with other isolated, interacting and merging systems from the literature shows that the galaxies studied in this paper are among the brightest interacting systems at X-ray, FIR, B-band and radio luminosities, and are even comparable to some late-stage merging systems. They all follow the established linear correlations for starburst systems between $L_X$--$L_{FIR}$, $L_X$--$L_{B}$ and $L_X$--$L_{1.4GHz}$, demonstrating that their X-ray output is dominated by emission from the starburst components, including NGC~7771 which may host a LLAGN. These types of in-depth studies of interacting and merging systems at X-ray energies with \xmmn and \chandra can assist in our understanding of the evolution of galaxies as a whole, revealing whether such interactions increase the activity levels in galaxies in the form of pure starbursts, and whether interactions play a major role in supplying sufficient fuel to trigger AGN activity that may be obscured at other wavelengths.

\section*{Acknowledgments}

This work is based on observations obtained with {\it XMM-Newton}, an ESA science mission with instruments and contributions directly funded be ESA and NASA. We thank the anonymous referee for helpful comments, and we thank R. Kilgard for supplying the integrated X-ray luminosities for the comparison isolated spiral galaxy sample prior to publication. This research has made use of the NASA/IPAC Extragalactic Database (NED) which is operated by the Jet Propulsion Laboratory, California Institute of Technology, under contract with NASA. The second digitized sky survey was produced by the Space Telescope Science Institute, under Contract No. NAS 5-26555 with NASA.  LPJ is supported by a PPARC studentship. AZ acknowledges partial support from NASA LTSA grant NAG5-13056 and NASA grant GO2-3111X.

\label{lastpage}

{}


\begin{thebibliography}{}

\bibitem[\protect\citeauthoryear{Alonso-Herrero et al.}{1999}]{alonso99} Alonso-Herrero A., Ward M.~J., Aragon-Salamanca A., Zamorano J., 1999, MNRAS, 302, 561 

\bibitem[\protect\citeauthoryear{Barnes \& Hernquist}{1991}]{barnes91} Barnes J.~E., Hernquist L.~E., 1991, ApJ, 370, L65 

\bibitem[\protect\citeauthoryear{Barnes \& Hernquist}{1996}]{barnes96} Barnes J.~E., Hernquist L., 1996, ApJ, 471, 115 

\bibitem[\protect\citeauthoryear{Behar et al.}{2001}]{behar01} Behar E., Rasmussen A.~P., Griffiths R.~G., Dennerl K., Audard M., Aschenbach B., Brinkman A.~C., 2001, A\&A, 365, L242

\bibitem[\protect\citeauthoryear{Bergvall, Laurikainen, \& Aalto}{Bergvall et al.}{2003}]{bergvall03} Bergvall N., Laurikainen E., Aalto S., 2003, A\&A, 405, 31 

\bibitem[\protect\citeauthoryear{Bohlin, Savage, \& Drake}{Bohlin et al.}{1978}]{bohlin78} Bohlin R.~C., Savage B.~D., Drake J.~F., 1978, ApJ, 224, 132 

\bibitem[\protect\citeauthoryear{Brandt, Halpern, \& Iwasawa}{Brandt et al.}{1996}]{brandt96} Brandt W.~N., Halpern J.~P., Iwasawa K., 1996, MNRAS, 281, 687

\bibitem[\protect\citeauthoryear{Bushouse}{1987}]{bushouse87} Bushouse H.~A., 1987, ApJ, 320, 49  

\bibitem[\protect\citeauthoryear{Bushouse, Werner, \& Lamb}{Bushouse et al.}{1988}]{bushouse88} Bushouse H.~A., Werner M.~W., Lamb S.~A., 1988, ApJ, 335, 74 

\bibitem[\protect\citeauthoryear{Colbert \& Mushotzky}{1999}]{colbert99} Colbert E.~J.~M., Mushotzky R.~F., 1999, ApJ, 519, 89 

\bibitem[\protect\citeauthoryear{Colbert et al.}{2004}]{colbert04} Colbert E.~J.~M., Heckman T.~M., Ptak A.~F., Strickland D.~K., Weaver K.~A., 2004, ApJ, 602, 231 

\bibitem[\protect\citeauthoryear{Condon}{1987}]{condon87} Condon J.~J., 1987, ApJS, 65, 485 

\bibitem[\protect\citeauthoryear{Condon et al.}{1990}]{condon90} Condon J.~J., Helou G., Sanders D.~B., Soifer B.~T., 1990, ApJS, 73, 359 

\bibitem[\protect\citeauthoryear{Condon, Anderson, \& Helou}{Condon et al.}{1991}]{condon91} Condon J.~J., Anderson M.~L., Helou G., 1991, ApJ, 376, 95 

\bibitem[\protect\citeauthoryear{Condon et al.}{1998}]{condon98} Condon J.~J., Cotton W.~D., Greisen E.~W., Yin Q.~F., Perley R.~A., Taylor G.~B., Broderick J.~J., 1998, AJ, 115, 1693

\bibitem[\protect\citeauthoryear{Cutri et al.}{2003}]{cutri03} Cutri R.~M., et al., 2003, yCat, 2246, 0 

\bibitem[\protect\citeauthoryear{David, Jones, \& Forman}{David et al.}{1992}]{david92} David L.~P., Jones C., Forman W., 1992, ApJ, 388, 82   

\bibitem[\protect\citeauthoryear{Davies, Alonso-Herrero, \& Ward}{Davies et al.}{1997}]{davies97} Davies R.~I., Alonso-Herrero A., Ward M.~J., 1997, MNRAS, 291, 557

\bibitem[\protect\citeauthoryear{Della Ceca et al.}{2002}]{dellaceca02} Della Ceca R., et al., 2002, ApJ, 581, L9  

\bibitem[\protect\citeauthoryear{de Naray et al.}{2000}]{denaray00} de Naray P.~J., Brandt W.~N., Halpern J.~P., Iwasawa K., 2000, AJ, 119, 612 

\bibitem[\protect\citeauthoryear{de Vaucouleurs}{1991}]{devaucouleurs91} de Vaucouleurs G., de Vaucouleurs A., Corwin H.~G., Buta R.~J., Paturel G., Fouque P., 1991, Third Reference Catalogue of Bright Galaxies

\bibitem[\protect\citeauthoryear{Dickey \& Lockman}{1990}]{dickey90} Dickey J.~M., Lockman F.~J., 1990, ARA\&A, 28, 215 

\bibitem[\protect\citeauthoryear{Elvis, Soltan, \& Keel}{Elvis et al.}{1984}]{elvis84} Elvis M., Soltan A., Keel W.~C., 1984, ApJ, 283, 479  

\bibitem[\protect\citeauthoryear{Englmaier \& Gerhard}{1997}]{englmaier97} Englmaier P., Gerhard O., 1997, MNRAS, 287, 57 

\bibitem[\protect\citeauthoryear{Fabbiano, Kim, \& Trinchieri}{Fabbiano et al.}{1992}]{fabbiano92} Fabbiano G., Kim D.-W., Trinchieri G., 1992, ApJS, 80, 531 

\bibitem[\protect\citeauthoryear{Fabbiano, Schweizer, \& Mackie}{Fabbiano et al.}{1997}]{fabbiano97} Fabbiano G., Schweizer F., Mackie G., 1997, ApJ, 478, 542 

\bibitem[\protect\citeauthoryear{Fabbiano, Zezas, \& Murray}{Fabbiano et al.}{2001}]{fabbiano01} Fabbiano G., Zezas A., Murray S.~S., 2001, ApJ, 554, 1035

\bibitem[\protect\citeauthoryear{Fabbiano \& Shapley}{2002}]{fabbiano02} Fabbiano G., Shapley A., 2002, ApJ, 565, 908 

\bibitem[\protect\citeauthoryear{Fabian et al.}{1998}]{fabian98} Fabian A.~C., Barcons X., Almaini O., Iwasawa K., 1998, MNRAS, 297, L11 

\bibitem[\protect\citeauthoryear{Gao et al.}{2003}]{gao03} Gao Y., Wang Q.~D., Appleton P.~N., Lucas R.~A., 2003, ApJ, 596, L171 

\bibitem[\protect\citeauthoryear{Garcia}{1993}]{garcia93} Garcia A.~M., 1993, A\&AS, 100, 47 

\bibitem[\protect\citeauthoryear{Gilfanov, Grimm, \& Sunyaev}{Gilfanov et al.}{2004}]{gilfanov04} Gilfanov M., Grimm H.-J., Sunyaev R., 2004, MNRAS, 347, L57 

\bibitem[\protect\citeauthoryear{Goldader et al.}{1997a}]{goldader97a} Goldader J.~D., Joseph R.~D., Doyon R., Sanders D.~B., 1997a, ApJS, 108, 449 

\bibitem[\protect\citeauthoryear{Goldader et al.}{1997b}]{goldader97b} Goldader J.~D., Joseph R.~D., Doyon R., Sanders D.~B., 1997b, ApJ, 474, 104

\bibitem[\protect\citeauthoryear{Guainazzi et al.}{2000}]{guainazzi00} Guainazzi M., Matt G., Brandt W.~N., Antonelli L.~A., Barr P., Bassani L., 2000, A\&A, 356, 463 

\bibitem[\protect\citeauthoryear{Helou, Soifer, \& Rowan-Robinson}{Helou et al.}{1985}]{helou85} Helou G., Soifer B.~T., Rowan-Robinson M., 1985, ApJ, 298, L7 

\bibitem[\protect\citeauthoryear{Henriksen \& Cousineau}{1999}]{henriksen99} Henriksen M., Cousineau S., 1999, ApJ, 511, 595

\bibitem[\protect\citeauthoryear{Hibbard \& van Gorkom}{1996}]{hibbard96} Hibbard J.~E., van Gorkom J.~H., 1996, AJ, 111, 655 

\bibitem[\protect\citeauthoryear{Ho, Filippenko, \& Sargent}{Ho et al.}{1997a}]{ho97a} Ho L.~C., Filippenko A.~V., Sargent W.~L.~W., 1997a, ApJ, 487, 591 

\bibitem[\protect\citeauthoryear{Ho, Filippenko, \& Sargent}{Ho et al.}{1997b}]{ho97b} Ho L.~C., Filippenko A.~V., Sargent W.~L.~W., 1997b, ApJS, 112, 315 

\bibitem[\protect\citeauthoryear{Ho et al.}{1997}]{ho97c} Ho L.~C., Filippenko A.~V., Sargent W.~L.~W., Peng C.~Y., 1997, ApJS, 112, 391 

\bibitem[\protect\citeauthoryear{Hummel et al.}{1990}]{hummel90} Hummel E., van der Hulst J.~M., Kennicutt R.~C., Keel W.~C., 1990, A\&A, 236, 333 

\bibitem[\protect\citeauthoryear{Jenkins et al.}{2004a}]{jenkins04a} Jenkins L.~P., Roberts T.~P., Warwick R.~S., Kilgard R.~E., Ward M.~J., 2004a, MNRAS, 349, 404 

\bibitem[\protect\citeauthoryear{Jenkins et al.}{2004b}]{jenkins04b} Jenkins L.~P., Roberts T.~P., Ward M.~J., Zezas A., 2004b, MNRAS, 352, 1335

\bibitem[\protect\citeauthoryear{Jim{\' e}nez-Bail{\' o}n et al.}{2003}]{jimenez03} Jim{\' e}nez-Bail{\' o}n E., Santos-Lle{\' o} M., Mas-Hesse J.~M., Guainazzi M., Colina L., Cervi{\~ n}o M., Gonz{\' a}lez Delgado R.~M., 2003, ApJ, 593, 127

\bibitem[\protect\citeauthoryear{Jogee}{2004}]{jogee04} Jogee S., 2004, LNP Volume on "AGN Physics on All Scales", Chapter 6, in press, preprint (astro-ph/0408383) 

\bibitem[\protect\citeauthoryear{Joseph et al.}{1984}]{joseph84} Joseph R.~D., Meikle W.~P.~S., Robertson N.~A., Wright G.~S., 1984, MNRAS, 209, 111

\bibitem[\protect\citeauthoryear{Karachentsev}{1972}]{karachentsev72} Karachentsev I.~D., 1972, SoSAO, 7, 1 

\bibitem[\protect\citeauthoryear{Karachentsev, Karachentsev, \& Lebedev}{Karachentsev et al.}{1988}]{karachentsev88} Karachentsev V.~E., Karachentsev I.~D., Lebedev V.~S., 1988, BuONC, 26, 37 

\bibitem[\protect\citeauthoryear{Keel et al.}{1985}]{keel85} Keel W.~C., Kennicutt R.~C., Hummel E., van der Hulst J.~M., 1985, AJ, 90, 708

\bibitem[\protect\citeauthoryear{Kennicutt \& Keel}{1984}]{kennicutt84} Kennicutt R.~C., Keel W.~C., 1984, ApJ, 279, L5 

\bibitem[\protect\citeauthoryear{Kennicutt et al.}{1987}]{kennicutt87} Kennicutt R.~C., Roettiger K.~A., Keel W.~C., van der Hulst J.~M., Hummel E., 1987, AJ, 93, 1011 

\bibitem[\protect\citeauthoryear{Kilgard et al.}{2002}]{kilgard02} Kilgard R.~E., Kaaret P., Krauss M.~I., Prestwich A.~H., Raley M.~T., Zezas A., 2002, ApJ, 573, 138

\bibitem[\protect\citeauthoryear{Kilgard et al.}{2004}]{kilgard04} Kilgard R.~E., Cowan J.~J., Garcia M.~R., et al., 2004, submitted to ApJS 

\bibitem[\protect\citeauthoryear{King}{2004}]{king04} King A.~R., 2004, MNRAS, 347, L18

\bibitem[\protect\citeauthoryear{Knapen, Shlosman, \& Peletier}{Knapen et al.}{2000}]{knapen00} Knapen J.~H., Shlosman I., Peletier R.~F., 2000, ApJ, 529, 93   

\bibitem[\protect\citeauthoryear{Knapen, P{\' e}rez-Ram{\'{\i}}rez, \& Laine}{Knapen et al.}{2002}]{knapen02} Knapen J.~H., P{\' e}rez-Ram{\'{\i}}rez D., Laine S., 2002, MNRAS, 337, 808 

\bibitem[\protect\citeauthoryear{Larson \& Tinsley}{1978}]{larson78} Larson R.~B., Tinsley B.~M., 1978, ApJ, 219, 46 

\bibitem[\protect\citeauthoryear{Lauberts}{1982}]{lauberts82} Lauberts A., 1982, Garching: European Southern Observatory (ESO)

\bibitem[\protect\citeauthoryear{Laurikainen, Salo, \& Buta}{Laurikainen et al.}{2004}]{laurikainen04} Laurikainen E., Salo H., Buta R., 2004, ApJ, 607, 103 

\bibitem[\protect\citeauthoryear{Lira et al.}{2002}]{lira02} Lira P., Ward M., Zezas A., Alonso-Herrero A., Ueno S., 2002, MNRAS, 330, 259

\bibitem[\protect\citeauthoryear{Lonsdale, Persson, \& Matthews}{Lonsdale et al.}{1984}]{lonsdale84} Lonsdale C.~J., Persson S.~E., Matthews K., 1984, ApJ, 287, 95 

\bibitem[\protect\citeauthoryear{Maiolino, Risaliti, \& Salvati}{Maiolino et al.}{1999}]{maiolino99} Maiolino R., Risaliti G., Salvati M., 1999, A\&A, 341, L35

\bibitem[\protect\citeauthoryear{Maiolino et al.}{2003}]{maiolino03} Maiolino R., et al., 2003, MNRAS, 344, L59  

\bibitem[\protect\citeauthoryear{Makishima et al.}{1986}]{makishima86} Makishima K., Maejima Y., Mitsuda K., Bradt H.~V., Remillard R.~A., Tuohy I.~R., Hoshi R., Nakagawa M., 1986, ApJ,  308, 635 

\bibitem[\protect\citeauthoryear{Makishima et al.}{2000}]{makishima00} Makishima K., et al., 2000, ApJ, 535, 632 

\bibitem[\protect\citeauthoryear{Matsumoto et al.}{2001}]{matsumoto01} Matsumoto H., Tsuru T.~G., Koyama K., Awaki H., Canizares C.~R., Kawai N., Matsushita S., Kawabe R., 2001, ApJ, 547, L25

\bibitem[\protect\citeauthoryear{McDowell et al.}{2003}]{mcdowell03} McDowell J.~C., et al., 2003, ApJ, 591, 154 

\bibitem[\protect\citeauthoryear{Miller et al.}{2003}]{miller03} Miller J.~M., Fabbiano G., Miller M.~C., Fabian A.~C., 2003, ApJ, 585, L37

\bibitem[\protect\citeauthoryear{Mitsuda et al.}{1984}]{mitsuda84} Mitsuda K., Inoue H., Koyama K., et al., 1984, PASJ,  36, 741

\bibitem[\protect\citeauthoryear{Mulchaey \& Regan}{1997}]{mulchaey97} Mulchaey J.~S., Regan M.~W., 1997, ApJ, 482, L135 

\bibitem[\protect\citeauthoryear{Nandra \& Pounds}{1994}]{nandra94} Nandra K., Pounds K.~A., 1994, MNRAS, 268, 405 

\bibitem[\protect\citeauthoryear{Neff \& Hutchings}{1992}]{neff92} Neff S.~G., Hutchings J.~B., 1992, AJ, 103, 1746 

\bibitem[\protect\citeauthoryear{Noguchi}{1988}]{noguchi88} Noguchi M., 1988, A\&A, 203, 259 

\bibitem[\protect\citeauthoryear{Nolan et al.}{2004}]{nolan04} Nolan L.~A., Ponman T.~J., Read A.~M., Schweizer F., 2004, MNRAS, 353, 221 

\bibitem[\protect\citeauthoryear{Nordgren et al.}{1997}]{nordgren97} Nordgren T.~E., Chengalur J.~N., Salpeter E.~E., Terzian Y., 1997, AJ, 114, 77 

\bibitem[\protect\citeauthoryear{Ott, Martin, \& Walter}{Ott et al.}{2003}]{ott03} Ott J., Martin C.~L., Walter F., 2003, ApJ, 594, 776 

\bibitem[\protect\citeauthoryear{Page et al.}{2004}]{page04} Page K.~L., O'Brien P.~T., Reeves J.~N., Turner M.~J.~L., 2004, MNRAS, 347, 316 

\bibitem[\protect\citeauthoryear{Persic \& Rephaeli}{2002}]{persic02} Persic M., Rephaeli Y., 2002, A\&A, 382, 843 

\bibitem[\protect\citeauthoryear{Pietsch et al.}{2001}]{pietsch01} Pietsch W.~et al., 2001, A\&A, 365, L174 

\bibitem[\protect\citeauthoryear{Ptak et al.}{1997}]{ptak97} Ptak A., Serlemitsos P., Yaqoob T., Mushotzky R., Tsuru T., 1997, AJ, 113, 1286

\bibitem[\protect\citeauthoryear{Ranalli, Comastri, \& Setti}{Ranalli et al.}{2003}]{ranalli03} Ranalli P., Comastri A., Setti G., 2003, A\&A, 399, 39  

\bibitem[\protect\citeauthoryear{Read}{2003}]{read03} Read A.~M., 2003, MNRAS, 342, 715 

\bibitem[\protect\citeauthoryear{Read \& Ponman}{1998}]{read98} Read A.~M., Ponman T.~J., 1998, MNRAS, 297, 143 

\bibitem[\protect\citeauthoryear{Rephaeli, Gruber, \& Persic}{Rephaeli et al.}{1995}]{rephaeli95} Rephaeli Y., Gruber D., Persic M., 1995, A\&A, 300, 91 

\bibitem[\protect\citeauthoryear{Roberts et al.}{2002}]{roberts02} Roberts T.~P., Warwick R.~S., Ward M.~J., Murray S.~S., 2002, MNRAS, 337, 677

\bibitem[\protect\citeauthoryear{Roberts et al.}{2004}]{roberts04} Roberts T.~P., Warwick R.~S., Ward M.~J., Goad M.~R., 2004, MNRAS, 349, 1193 

\bibitem[\protect\citeauthoryear{Sanders \& Mirabel}{1996}]{sanders96} Sanders D.~B., Mirabel I.~F., 1996, ARA\&A, 34, 749

\bibitem[\protect\citeauthoryear{Sanders et al.}{2003}]{sanders03} Sanders D.~B., Mazzarella J.~M., Kim D.-C., Surace J.~A., Soifer B.~T., 2003, AJ, 126, 1607 

\bibitem[\protect\citeauthoryear{Schmitt}{2001}]{schmitt01} Schmitt H.~R., 2001, AJ, 122, 2243 

\bibitem[\protect\citeauthoryear{Smith et al.}{1999}]{smith99} Smith D.~A., Herter T., Haynes M.~P., Neff S.~G., 1999, ApJ, 510, 669 

\bibitem[\protect\citeauthoryear{Stevens, Read, \& Bravo-Guerrero}{Stevens et al.}{2003}]{stevens03} Stevens I.~R., Read A.~M., Bravo-Guerrero J., 2003, MNRAS, 343, L47 

\bibitem[\protect\citeauthoryear{Strickland \& Stevens}{2000}]{stricklandstevens00} Strickland D.~K., Stevens I.~R., 2000, MNRAS, 314, 511

\bibitem[\protect\citeauthoryear{Strickland et al.}{2000}]{strickland00a} Strickland D.~K., Heckman T.~M., Weaver K.~A., Dahlem M., 2000, AJ, 120, 2965 

\bibitem[\protect\citeauthoryear{Strickland et al.}{2002}]{strickland02} Strickland D.~K., Heckman T.~M., Weaver K.~A., Hoopes C.~G., Dahlem M., 2002, ApJ, 568, 689

\bibitem[\protect\citeauthoryear{Surace, Sanders, \& Mazzarella}{Surace et al.}{2004}]{surace04} Surace J.~A., Sanders D.~B., Mazzarella J.~M., 2004, AJ, 127, 3235

\bibitem[\protect\citeauthoryear{Swartz et al.}{2003}]{swartz03} Swartz D.~A., Ghosh K.~K., McCollough M.~L., Pannuti T.~G., Tennant A.~F., Wu K., 2003, ApJS, 144, 213  

\bibitem[\protect\citeauthoryear{Toomre \& Toomre}{1972}]{toomre72} Toomre A., Toomre J., 1972, ApJ, 178, 623

\bibitem[\protect\citeauthoryear{Tsch{\" o}ke, Hensler, \& Junkes}{Tsch{\" o}ke et al.}{2000}]{tschoke00} Tsch{\" o}ke D., Hensler G., Junkes N., 2000, A\&A, 360, 447

\bibitem[\protect\citeauthoryear{Tully}{1988}]{tully88} Tully R.~B., 1988, Nearby Galaxies Catalogue, Cambridge University Press

\bibitem[\protect\citeauthoryear{Veilleux et al.}{1995}]{veilleux95} Veilleux S., Kim D.-C., Sanders D.~B., Mazzarella J.~M., Soifer B.~T., 1995, ApJS, 98, 171 

\bibitem[\protect\citeauthoryear{Wolter \& Trinchieri}{2004}]{wolter04} Wolter A., Trinchieri G., 2004, A\&A, in press, preprint (astro-ph/0407446)

\bibitem[\protect\citeauthoryear{Zezas, Georgantopoulos, \& Ward}{Zezas et al.}{1998}]{zezas98} Zezas A.~L., Georgantopoulos I., Ward M.~J., 1998, MNRAS, 301, 915

\bibitem[\protect\citeauthoryear{Zezas, Ward, \& Murray}{Zezas et al.}{2003}]{zezas03} Zezas A., Ward M.~J., Murray S.~S., 2003, ApJ, 594, L31

\bibitem[\protect\citeauthoryear{Zezas et al.}{2002}]{zezas02} Zezas A., Fabbiano G., Rots A.~H., Murray S.~S., 2002, ApJS, 142, 239 


\end{thebibliography}
\end{document}